\documentclass[conference]{IEEEtran}
\ifCLASSINFOpdf
\else
\fi

\usepackage{tikz}
\usepackage{amsmath}
\usepackage{url}
\usepackage{filecontents}
\usepackage{verbatim}
\usepackage{amsmath,amssymb,amsfonts}
\usepackage{graphicx}
\usepackage{textcomp}
\usepackage{xcolor}
\usepackage{subfigure} 
\usepackage{breakurl}
\usepackage{multirow}
\usepackage{diagbox}
\usepackage{makecell}
\usepackage{booktabs}
\usepackage[normalem]{ulem}
\let\labelindent\relax
\usepackage{enumitem}
\usepackage{fancyhdr} 
\usepackage{booktabs}
\usepackage{tikz}
\usepackage{filecontents}
\usepackage{colortbl}
\usepackage{balance}
\usepackage{CJKutf8}
\usepackage{threeparttable}
\usepackage{bbding}
\usepackage{hyperref} 

\usepackage{longtable}
\usepackage{supertabular,booktabs}
\newcommand{\xw}[1]{\textcolor{black}{#1}}

\newcommand{\Major}[1]{\textcolor{black}{#1}}

\newcommand{\NewMajor}[1]{\textcolor{black}{#1}}

\usepackage{xcolor,pifont}
\newcommand*\colourcheck[1]{%
  \expandafter\newcommand\csname #1check\endcsname{\textcolor{#1}{\ding{52}}}%
}
\newcommand*\colourcross[1]{%
  \expandafter\newcommand\csname #1cross\endcsname{\textcolor{#1}{\ding{56}}}%
}
\colourcheck{black}
\colourcheck{red}

\colourcross{black}
\colourcross{green}

\newcommand{\tabincell}[2]{\begin{tabular}{@{}#1@{}}#2\end{tabular}}

\begin{document}
\title{Exploiting Sequence Number Leakage: TCP Hijacking in NAT-Enabled Wi-Fi Networks}

\author{
    \IEEEauthorblockN{Yuxiang Yang\IEEEauthorrefmark{1}, Xuewei Feng\IEEEauthorrefmark{1}, Qi Li\IEEEauthorrefmark{2}\IEEEauthorrefmark{4}, Kun Sun\IEEEauthorrefmark{3}, Ziqiang Wang\IEEEauthorrefmark{5}, and Ke Xu\IEEEauthorrefmark{1}\IEEEauthorrefmark{4}\Envelope}
    
    \IEEEauthorblockA{\IEEEauthorrefmark{1}Department of Computer Science and Technology \& BNRist, Tsinghua University}
    
    \IEEEauthorblockA{\IEEEauthorrefmark{2}Institute for Network Sciences and Cyberspace \& BNRist, Tsinghua University, \IEEEauthorrefmark{4}Zhongguancun Lab}
    
    \IEEEauthorblockA{\IEEEauthorrefmark{3}Department of Information Sciences and Technology \& CSIS, George Mason University}
    
    \IEEEauthorblockA{\IEEEauthorrefmark{5}School of Cyber Science and Engineering, Southeast University}
    %\IEEEauthorblockA{\IEEEauthorrefmark{4}Zhongguancun Lab}
   
   \{yangyx22@mails, qli01@, xuke@\}tsinghua.edu.cn, fengxw06@126.com, ksun3@gmu.edu, ziqiangwang@seu.edu.cn \\
}

\iffalse
\author{\IEEEauthorblockN{Yuxiang Yang}
\IEEEauthorblockA{Tsinghua University\\
yangyx22@mails.tsinghua.edu.cn}
\and
\IEEEauthorblockN{Xuewei Feng}
\IEEEauthorblockA{Tsinghua University\\
brafum@yeah.net}
\and
\IEEEauthorblockN{Qi Li}
\IEEEauthorblockA{Tsinghua University\\
someemail@somedomain.com}
}
\fi


%Telephone: (800) 555--1212, Fax: (888) 555--1212}


\IEEEoverridecommandlockouts
\makeatletter\def\@IEEEpubidpullup{6.5\baselineskip}\makeatother
\IEEEpubid{\parbox{\columnwidth}{
    Network and Distributed System Security (NDSS) Symposium 2024\\
    26 February - 1 March 2024, San Diego, CA, USA\\
    ISBN 1-891562-93-2\\
    https://dx.doi.org/10.14722/ndss.2024.23419\\
    www.ndss-symposium.org
}
\hspace{\columnsep}\makebox[\columnwidth]{}}

\maketitle

\begin{abstract}
% In this paper, we uncover a subtle yet serious side channel 
In this paper, we uncover a \Major{new} side-channel vulnerability in the widely used NAT port preservation strategy and an insufficient reverse path validation strategy of Wi-Fi routers, which allows an off-path attacker to infer if there is one victim client in the same network communicating with another host on the Internet using TCP. After detecting the presence of TCP connections between the victim client and the server, the attacker can evict the original NAT mapping and reconstruct a new mapping at the router by sending fake TCP packets due to the routers' vulnerability of disabling TCP window tracking strategy, which has been faithfully implemented in most of the routers for years. In this way, the attacker can intercept TCP packets from the server and obtain the current sequence and acknowledgment numbers, which in turn allows the attacker to \Major{forcibly close} the connection, poison the traffic in plain text, or reroute the server's incoming packets to the attacker.

% We tested 59 popular routers from 22 well-known vendors 
We test \Major{67 widely used} routers from \Major{30} vendors and discover that \Major{52} of them are affected by this attack. Also, we conduct an extensive measurement study on 93 real-world Wi-Fi networks. The experimental results show that 75 of these evaluated Wi-Fi networks (81\%) are \Major{fully} vulnerable to our attack. \Major{Our case study shows that it takes about 17.5, 19.4, and 54.5 seconds on average to terminate an SSH connection, download private files from FTP servers, and inject fake HTTP response packets with success rates of 87.4\%, 82.6\%, and 76.1\%}. We responsibly disclose the vulnerability and suggest mitigation strategies to all affected vendors and have received positive feedback, including acknowledgments, CVEs, rewards, and adoption of our suggestions.

\end{abstract}


% \fi
%%\IEEEpeerreviewmaketitle

%-------------------------------------------------------------------------------
\section{Introduction}
%-------------------------------------------------------------------------------
%Wi-Fi has become a ubiquitous technology, revolutionizing the way we connect to the Internet. With the increasing demand for connectivity, Wi-Fi has emerged as one of the most popular technology to provide wireless Internet access. 
%With the growing demand for connectivity, 
Wi-Fi has emerged as one of the most popular technologies for providing Internet access, being widely used in restaurants, offices, coffee shops, airports, and other public places.
%
%However, open Wi-Fi networks are vulnerable to a range of attacks including eavesdropping, packet sniffing, and man-in-the-middle (MitM) attacks from malicious actors, which can result in compromised user data, identity theft, and other forms of malicious activity.
%
% As Wi-Fi has become an essential part of our daily lives, it is increasingly important to take measures to secure these networks. Nowadays with the widespread deployment of wireless security mechanisms (e.g., WPA2 and WPA3), it is difficult for an off-path attacker (i.e., with no control over the wireless router) to obtain the communication information between other clients in the same Wi-Fi network and outside servers, even if it can sniff frames in the wireless channels as those frames are always encrypted. As a result, it is believed that Wi-Fi networks are secure from traffic hijacking, especially considering that security vulnerabilities reported in recent years have been addressed~\cite{wifi-protect-1, Vanhoef21Fragment, Vanhoef20dragonblood}).
%
\Major{However, Wi-Fi networks are often exploited by malicious attackers to launch various attacks. In addition to exploiting vulnerabilities to break the protection of encryption~\cite{vanhoef2017key, vanhoef2018release, Vanhoef20dragonblood}, a lot of prior works have already been conducted on session hijacking in Wi-Fi networks~\cite{schepers2023framing,Vanhoef21Fragment,Tolley21VPN,rogueAP-attack,arp-attack}, 
% Most previous attacks relied on the flaws of the protocols
e.g., injecting forged wireless frames via vulnerabilities in WPA2 implementations~\cite{schepers2023framing,Vanhoef21Fragment}, eavesdropping on wireless channels~\cite{ohigashi2009practical}, intercepting packets via side channels in VPN tunnels of wireless routers~\cite{Tolley21VPN}, creating a rogue clone (i.e., Evil-Twin) of the network~\cite{rogueAP-attack}, or just abusing the classic ARP poisoning attack~\cite{arp-attack} to hijack the communication between victim clients and servers, thus disrupting normal user usage, stealing confidential information, and potentially causing financial losses. 
% At the same time, the evolution of attack technology continues to promote the advancement of defense technology. 
Fortunately, most of the prior attacks have been repaired or mitigated and targeted defense measures have been proposed as well~\cite{wifi-protect-1, arp-1,Vanhoef21Fragment, Vanhoef20dragonblood}. Nowadays, with the widespread deployment of wireless security mechanisms (e.g., WPA2 and WPA3) and the adoption of protection strategies (e.g., AP isolation, ARP prevention, and Rogue AP detection), it is increasingly difficult for an off-path attacker (i.e., with no control over the router) to obtain the communication information between other clients in the same Wi-Fi network and outside servers.
}

In public Wi-Fi networks, network address translation (NAT) is widely used to save IPv4 address space and protect internal clients from being identified by external attackers.
After attaching to the same Wi-Fi network enabling NAT, clients share the external IP address to access the Internet. When it takes the upper protocols (e.g., TCP and UDP) into consideration, the router will create NAT mappings to keep track of the connections, which record the IP addresses, upper-level information such as protocol, ports, timeout, and reply status, etc. 
In most cases, the router tries to keep the layer-4 information the same as the originators, such as the TCP source port, which is the so-called \textit{port preservation} strategy~\cite{rfc5382}. However, cases are that some clients in the LAN may communicate with the same remote server with the same source port at the same time as they have no idea about each other. Although with very little probability, the router has to deal with these cases and it will assign a new TCP source port, change the IP address, and port at the same time when TCP packets pass through it. Besides, due to reasons such as performance considerations, the router will not record all of the session information in the NAT mappings, such as tracking the current TCP window. Thus, it will not check the sequence and acknowledgment numbers strictly when TCP packets arrive.

In this paper, we uncover a new off-path TCP hijacking attack in Wi-Fi networks that exploits vulnerabilities in the NAT mapping strategies of routers. The attack includes three steps. 
% %
% First, the attacker scans the network to find potential victims attached to the same access point (AP) and probes the router's external IP address. 
% %
% 
%
First, the attacker probes the router's external IP address, identifies whether AP isolation is enabled and scans to find potential victims in the same network \Major{when it is disabled}.
Second, the attacker infers the presence of TCP connections between any client and a remote server by sending fake TCP \texttt{SYN} and \texttt{SYN/ACK} packets.
Third, the attacker evicts the original NAT mapping of the victim connection with forged \texttt{RST} packets and replaces it with a new mapping at the router by sending a TCP data packet to the server. 
After that, it can intercept the \texttt{ACK} packet from the server that is meant to send to the victim and thus obtain the sequence and acknowledgment numbers within it so as to completely hijack the TCP connection.
The attacker only needs to connect to the same \Major{network} as the victim client, and it does not need any assistance of malicious puppets, i.e., unprivileged applications or sandboxed scripts deployed on victim clients. \Major{Compared with prior attacks, our work sheds light on the vulnerabilities existing in the abusing peculiarities of NAT strategies and behaviors of routers instead of flaws in TCP specifications,
and our attack is not limited to specific scenarios or applications (e.g., WPA2/WPA3, or VPNs).} Besides, the OS types or versions of the clients and servers are unrestricted in our attack \Major{in contrast to previous TCP hijacking attacks that can only target servers or clients with specific operating systems}\cite{yossi-tcp,qian2012,qian2012-2}.

%As all clients access the Internet through the same router, they will share the resources of the router inevitably, which provides the attacker with the possibility of inferring other clients' network communications. As 
In our investigations, most Wi-Fi routers (e.g., Asus, Netgear, Linksys, TP-Link, Huawei, and Xiaomi) adopt the \textit{port preservation} strategy when creating new NAT mappings for TCP connections initiated by internal clients~\cite{Amit-devicetracking,tpds-Herzberg}. The attacker can intentionally initiate a connection, i.e., sending a \texttt{SYN} packet, to the target server with a guessed client's port and distinguish the guess by observing whether the port will be changed at the router as a collision will happen if it is a right guess. The attacker can send a spoofed \texttt{SYN/ACK} packet with a source address of the remote server, a destination address of the external IP of the router, and a destination port of the guessed port as a response to verify if the port is changed. If the router disobeys the RFC recommendation to enable the reverse path validation with a strict mode \cite{rfc2827,rfc3704}, the forged \texttt{SYN/ACK} packet cannot be detected and will not be dropped by the router, which is often the case in most routers we tested. If there is any connection from the LAN to a target remote server with the guessed source port, the router will choose another source port to initiate the connection, and then the \texttt{SYN/ACK} packet will be forwarded to the victim. Yet if there is no connection with this source port from the LAN to the server, the router will keep the port to initiate the connection, and then the \texttt{SYN/ACK} packet will be forwarded back to the attacker. In this way, the attacker can infer whether any client is communicating with the server and the source port of the client if there is such a TCP connection.

%We have found another vulnerability of the TCP window tracking strategy in routers, which can be abused by an off-path attacker to obtain the sequence and acknowledgment numbers of the connection once the source port is known. 
After identifying a target TCP connection, the attacker can directly get the sequence and acknowledgment numbers of the connection by exploiting a new vulnerability arising in the disabled TCP window tracking strategy of Wi-Fi routers.
As routers pursue higher performance, they choose to disable TCP window tracking by default, i.e., they will not check the sequence and acknowledgment numbers strictly in TCP packets. So the attacker can send forged TCP reset packets to clean the NAT mapping of the victim connection. \Major{After waiting for the timeout of the NAT mapping (i.e., 1 second or 10 seconds)}, the attacker can send a TCP data packet using its private IP address and the same source port to the server with arbitrary sequence and acknowledgment numbers. The router will only translate the IP address of the packet except for the source port, as there is no port collision anymore. And the packet will match the victim connection from the perspective of the server, which will return a TCP \texttt{ACK} packet carrying the exact sequence and acknowledgment numbers of the victim connection upon seeing the packet with wrong numbers~\cite{rfc793}. When arriving at the router, this \texttt{ACK} packet will be routed to the attacker as the NAT mapping has been falsified, and thus it steals the sequence and acknowledgment numbers of the victim connection easily, i.e., without traversing the 32-bit space to infer these numbers as previous methods \cite{cao2016off,ccsfeng,chen2018off}. 
\Major{It should be noted that the NAT mapping timeout will be refreshed if there are related packets traveling through the router, which may interfere with the attack. We will analyze the detailed influence in Section \ref{sec:practical_consideration}}.

% Once the sequence and acknowledgment numbers are obtained by the attacker, it can choose to terminate the victim TCP connection directly by sending crafted \texttt{RST} packets or to poison the victim TCP traffic by sending forged data packets after restoring the mapping for the victim client via issuing spoofed TCP \texttt{RST} and \texttt{ACK} packets.
% %
% Note that traffic encryption (e.g., HTTPS) may disturb the attacker's poisoning.
% However, there are still about 20\% of websites that transmit traffic in plaintext according to the reports on HTTPS adoption\footnote{See \url{https://w3techs.com/technologies/details/ce-httpsdefault} for daily statistics on HTTPS adoption.}.
% % 
% Alternatively, the attacker can also keep the NAT mapping and replaces the victim by itself to act as a MITM, since the router will continue forwarding packets (intended for the victim client) to the attacker instead. 

Once the sequence and acknowledgment numbers are obtained by the attacker, it can choose to \Major{launch three types of attacks: (i) \textbf{TCP Denial-of-Service (DoS) attack} to terminate victim TCP connections directly by sending \texttt{RST} packets. 
(ii) \textbf{TCP hijacking attack} to take over the NAT mapping and replace the victim by itself since the router will continue forwarding packets (intended for the victim client) to the attacker instead.
(iii) \textbf{TCP injection attack} to poison the victim TCP traffic by sending crafted data packets after restoring the mapping for the victim client via issuing spoofed TCP \texttt{RST} and \texttt{ACK} packets.
Note that traffic encryption (e.g., HTTPS) may disturb the attacker's poisoning.
However, about 20\% of websites still transmit traffic in plaintext according to the reports on HTTPS adoption\footnote{See \url{https://w3techs.com/technologies/details/ce-httpsdefault} for daily statistics on HTTPS adoption.}. AP isolation may also influence the TCP injection attack due to the requirement of reconstructing the client's original NAT mapping. However, the other two attacks (i.e., TCP DoS and hijacking attacks) are not affected, which will be illustrated in Section \ref{sec:practical_consideration}.
}
We conduct a large-scale empirical study to demonstrate that the attack can be performed to cause \Major{potential} damage in the real world. 
First, we investigate the default settings of routers on the market and have tested \Major{67 widely used} router models from \Major{30} vendors and find that \Major{52} of them from \Major{24} vendors are vulnerable to the attack.  
% Moreover, our extensive measurement results show 
Moreover, our \Major{empirical} measurement results show that the attacks can be successfully performed in various real-world Wi-Fi networks. 
We evaluate 93 Wi-Fi networks in six months, including most of the popular Wi-Fi scenarios (e.g., Wi-Fi networks in coffee shops, hotels, bookstores, and enterprises). The experimental results show that 75 (81\%) out of these evaluated Wi-Fi networks are fully vulnerable to our attacks. We implement a PoC and perform case studies on applications like \Major{SSH, FTP, and HTTP} to validate the effectiveness of the attack. In our experiments, an off-path attacker can detect and terminate an SSH connection in 17.5 seconds with a success rate of 87.4\%, \Major{download private files from an FTP server within 19.4 seconds with a success rate of 82.6\%}, and manipulate web traffic within 54.5 seconds with a success rate of 76.1\%, on average. These results demonstrate that this attack is feasible and may throw \Major{potential} threats to normal Wi-Fi users.

Finally, we identify the root cause and suggest mitigation to void this attack with the intuitive idea of breaking the conditions of the attack. Besides, we have responsibly disclosed the vulnerability to the affected router vendors and the OpenWrt community with affirmative feedback. At the time of writing, researchers from the OpenWrt community and \Major{7} of these vendors have confirmed the vulnerability and are repairing it in their products according to our suggestions. In addition, 10 CVE numbers have been assigned for this vulnerability from different vendors
% (i.e., from CVE-2023-30305 to CVE-2023-30314)
(i.e., from CVE-2023-30305 to CVE-2023-30314). 
The rest vendors are still in the process of investigating the vulnerability.

\noindent \textbf{Contributions}. Our main contributions are the following:

\begin{itemize}
%\begin{itemize}
%[leftmargin=*]

	% \item {We uncover a new side channel vulnerability of the NAT behaviors in Wi-Fi networks that can be exploited to hijack TCP connections by off-path attackers.}
 	\item {We uncover a new side channel vulnerability of the NAT behaviors in Wi-Fi networks that can be exploited to \Major{attack} TCP connections by off-path malicious insiders.}

	% \item {We uncover a new side channel in Wi-Fi networks and demonstrate that it can be exploited to learn the presence of others' TCP connections and intercept the sequence and acknowledgment numbers of the connections.}
	
	\item {We perform a large-scale measurement and reveal a number of routers vulnerable to the attack. Our extensive evaluations against \Major{67 widely used} router models and case studies in 93 various Wi-Fi networks show that our attacks can cause \Major{potential} damage in the real world.}
	
	\item {We suggest three countermeasures by eliminating the conditions to fight back the attack, and some of them have been adopted by the affected manufacturers.}

\end{itemize}

% \noindent \textbf{Paper organization.} The remainder of this paper is structured as follows. Section \ref{sec:background} provides an overview of the technical background of NAT and port allocation strategies, TCP window tracking in routers, and reverse path validation. Section \ref{sec:threat_model} describes our threat model and requirements for the attack. Section \ref{sec:attack_procedure} illustrates the steps of the attack procedure. Section \ref{sec:empirical_study} elaborates on our empirical study of vulnerable routers and case studies in various Wi-Fi networks. Section \ref{sec:discussion} discusses the factors impacting the effectiveness of the attack and the extension of our attack model. Section \ref{sec:countermeasures} shows the ethical disclosure process and proposes mitigation. Section \ref{sec:relatedwork} summarizes related work and Section \ref{sec:conclusion} concludes the paper.

%-------------------------------------------------------------------------------
\section{Background}\label{sec:background}
%-------------------------------------------------------------------------------
\subsection{NAT and Port Allocation Strategies}\label{sec:background-NAT}

Network Address Translation (NAT) is a technology developed to solve the shortage of IPv4 addresses and hide the network topology from an external entity \cite{nat05}, which is widely used by routers in Wi-Fi networks. 
When packets traverse through the router, it has to translate the IP addresses of the packets between internal and external addresses\footnote{Since our work considers multiple levels of NAT, the router's external IP address may not be a public IP address.} and record the other necessary information of the related connection. The router maintains a NAT mapping table to keep track of the internal IP addresses and ports associated with each corresponding external IP address and port, which allows incoming packets to be directed to the correct host on the private network. Since our work focuses on the TCP protocol, we will illustrate the NAT behavior of TCP mappings henceforth.

When an internal host initiates a connection to an external server, i.e., sending a \texttt{SYN} packet, the router will create a new mapping in the table, which is called a binding in NAT terminology \cite{rfc2663}. 
\Major{Besides, we find that not only \texttt{SYN} packets but also packets with  \texttt{PUSH}, or \texttt{ACK} flags can incur new NAT mappings at the router.}
The mapping will record the source IP addresses and ports translated before and after, the destination IP address and port, protocol, session state, and corresponding mapping timeout. 
% After the replies from the external host arrive, the router finds the corresponding mapping in the table, translates the IP address and port, and forwards the packet to the internal host, updating the state of the session mapping simultaneously.
%
After the replies from the external host arrive, the router forwards the packets to the internal host according to the mapping and updates its state simultaneously.
Since the related RFCs have not proposed a fixed strategy for the translation behavior of source ports, 
% the strategy to deal with the source port of TCP packets is different depending on the implementation of NAT devices, which includes the following strategies \cite{tpds-Herzberg}:
it can be different depending on the implementation of NAT devices, which includes the following strategies \cite{tpds-Herzberg}:

(1) \textit{port preservation}, where the NAT device attempts to preserve the source port if possible. When a collision happens, i.e., different internal hosts choose the same source port to communicate with the same external host of the same port, the NAT device should resolve the collision by selecting a new port (e.g., another random unused port).
(2) \textit{random selection}, where the NAT device translates the source port to another random port from a pool of available ports.
(3) \textit{sequential selection}, where the NAT device selects a random port for the first connection to each destination and translates the ports of subsequent packets to that destination consecutively based on the first port. 
(4) \textit{port overloading}, where the NAT device always uses \textit{port preservation} even in the case of collision. In this case, new connections will take over the original mapping, and the old connection will be disturbed, which is not recommended in RFC 5382 \cite{rfc5382}.

As with any stateful middle device, routers have to manage the state of mappings and track active flows. 
% The mappings will be released when they are no longer needed. 
Generally, the routers often rely on both the states of connections and timeouts of mappings to prune unnecessary NAT mappings. RFC 5382 recommends that the minimum timeout for the \texttt{ESTABLISHED} state is 2 hours and 4 minutes, which is faithfully implemented in most routers \cite{rfc5382}, and the routers also set timeouts for other states (e.g., 1 second or 10 seconds for the \texttt{CLOSE} state which the mapping will turn into upon seeing corresponding \texttt{RST} packets).

\subsection{TCP Window Tracking in Routers}\label{sec:tcp_window_tracking}

As a middle device between the client and server, the router has to record the connection information of the related hosts for subsequent packet delivery. However, as the TCP protocol was originally designed for end-to-end communication and did not take the middle devices into consideration, the router cannot and will not record all of the information due to many reasons (e.g., performance considerations). For instance, the router will choose not to track the current TCP window of the connection, and thus it will not check the sequence and acknowledgment numbers of TCP packets strictly. 
% The most famous open-sourced router operating systems, i.e., OpenWrt and AsusWrt, both have related options to reduce CPU overhead.
The open-sourced router operating systems, i.e., OpenWrt and AsusWrt, both have related options to reduce CPU overhead, i.e., the \texttt{nf\_conntrack\_tcp\_no\_window\_check} option in OpenWrt and the \texttt{ip\_conntrack\_tcp\_be\_liberal} option in AsusWrt. These options are set to true by default, and once they are set, Netfilter \cite{netfilter} will not perform TCP window tracking in contrast to the original Linux kernel. 
The difference between the two systems is that OpenWrt does not check the sequence number of the packet at all, while AsusWrt only checks if the sequence number is beyond the current sequence number in a 2G space.
Besides, we found most of the routers in the market also disable the TCP window tracking strategy by default and have similar behaviors to the two systems above.

We will show that routers disabling TCP window tracking can be abused by an off-path attacker to clean the NAT mappings of other clients with forged \texttt{RST} packets. 
For OpenWrt-based routers and those with similar settings, the attacker can use one forged \texttt{RST} packet with any arbitrary sequence number to clean the mapping, and for AsusWrt-based routers and those with similar settings, the attacker can forge two \texttt{RST} packets specified with two sequence numbers in the gap of 2G to bypass the range check easily and effectively.

\subsection{Reverse Path Validation}\label{sec:rp_validation}

% by checking by checking
To prevent IP spoofing attacks and promote the process of source address validation, RFC 2827 and RFC 3704 propose the concept of reverse path validation, which verifies the authenticity of inbound traffic by checking whether the source IP address can be routed back via the interface on which packets are received against the routing table, to ensure they come from an authorized sender \cite{rfc2827,rfc3704}. With this strategy enabled, only if the packets can be routable back from the incoming interface will they be processed by the kernel and routed to their destinations. Otherwise, they will be dropped. Most Linux-based systems control the strategy through the \textit{rp\_filter} kernel variable, which offers three options \cite{rpfilter}:

\begin{itemize}
\item {\textbf{0:} In this mode, the source address validation is disabled.}

\item {\textbf{1:} Strict Mode as defined in RFC3704. In this mode, the device should compare the source address of incoming packets to the Forwarding Information Base (FIB). If the incoming interface is not the best reverse path, packets will be dropped.}

\item{\textbf{2:} Loose Mode as defined in RFC3704. In this mode, the device compares the source address of incoming packets against the FIB, and only if the packets are not reachable via any interface will they be dropped.}
\end{itemize}
 
% Linux set its default value to 0. 
RFC 3704 recommends using the strict mode to prevent IP spoofing attacks. The loose mode is recommended if the device uses asymmetric routing (e.g., a mobile phone with a Wi-Fi interface and multiple interfaces for receiving packets from cell towers) or other complicated routing strategies. \Major{Previous research~\cite{Tolley21VPN} has shown that in the VPN scenarios, the lack of reverse path validation on client devices allows a blind in-path attacker (e.g., a router controlled by an attacker) to spoof packets to learn the virtual IP used by the tun0 interface for the VPN connection and infer the necessary fields to hijack the active connection.} By contrast, we find that most routers also do not obey the recommendation, and they will not drop packets with spoofed source addresses matching a connection in the NAT mappings and will accept them on any interface. 

% For example, questions are that why OpenWrt disables the reverse path validation in many forums. The answer is that OpenWrt inherits the settings from Linux. Then most OpenWrt-based routers are vulnerable to IP spoofing attacks \cite{rpfilter_forum}.

We will show that an off-path attacker in the LAN can abuse routers without reserve path validation to forward spoofed \texttt{SYN/ACK} packets with the server's IP address as the source and the router's external IP address as destination, which can be leveraged to infer source ports of connections used by other clients through observing the whereabouts of these \texttt{SYN/ACK} packets. Additionally, the attacker can also send forged \texttt{RST} packets to the router's external IP address. Though the source address specified in the packets is the server, the router without reverse path validation will process them in the kernel mistakenly and thus change the state of the NAT mappings to \texttt{CLOSE}, leading to our attack.

%-------------------------------------------------------------------------------
\section{Threat Model}\label{sec:threat_model}
\begin{figure}[h]
	\begin{center}
            \vspace{7pt}	
		\includegraphics[width=0.85\linewidth]{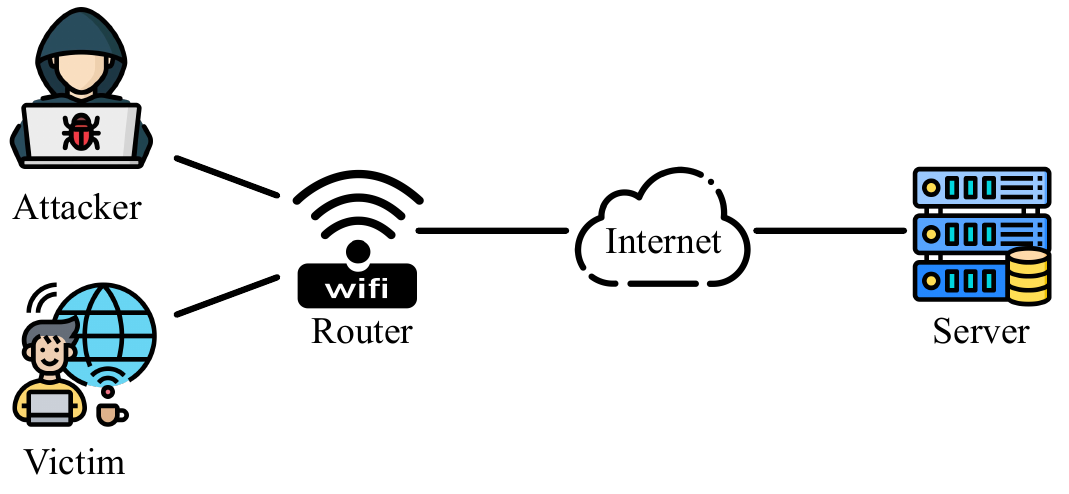}
		% \vspace{2mm}
		\caption{Threat model of TCP hijacking attacks in Wi-Fi networks.}
		\label{threat-model}
	\end{center}
	% \vspace{-4mm}
\end{figure}

Figure~\ref{threat-model} illustrates the threat model of our off-path TCP attacks in Wi-Fi networks. The model consists of three hosts and one router, namely, a remote server, a victim client, an off-path attacker, and a vulnerable router. The remote server may be a web application, an SSH \Major{or FTP} server in different attack scenarios. 
The victim client (e.g., a mobile phone or a laptop) is connected to a wireless access point to communicate with the remote server on the Internet, i.e., visiting web pages, downloading files through FTP, or using the SSH service to control remote hosts.
The off-path attacker is a malicious client who can access the same Wi-Fi network as the victim client. 
%Before launching the TCP hijacking attacks, it can infer the communication information between the victim client and the remote server. 
A router acts as the gateway of clients in the LAN to provide Internet services for the Wi-Fi network.
% however, the router has vulnerabilities in the NAT mapping strategies and does not deploy the reverse path validation to filter spoofed packets. 
% in public places (e.g., coffee shops, restaurants, hotels, and shopping malls). 

\NewMajor{
Existing studies~\cite{rogueAP-attack,evil-AP} demonstrate that a malicious insider can create an evil twin of the network and trick the victims into connecting to it by broadcasting the same SSID in the open (with no encryption) or home mode (accessed through pre-shared key) Wi-Fi networks, thus hijacking the traffic in the network. 
However, these attacks can be throttled by existing defenses, e.g., Rogue AP detection \cite{Huawei-rogueAP,Linksys-rogueAP}.  
It is widely believed that only AP isolation enabled enterprise mode Wi-Fi networks can effectively protect clients from each other, whereas open and home mode Wi-Fi networks face challenges in preventing insider threats. 
In this work, we propose a novel attack that can evade all defenses above in Wi-Fi networks. 
As a result, our attack holds particular significance for enterprise mode Wi-Fi networks, differentiating it from the rogue clone attacks in open and home mode Wi-Fi networks. Moreover, our attack can serve as an alternative method to compromise open and home mode Wi-Fi networks.
In our attack, we assume that with the deployment of security mechanisms (e.g., WPAs) and the usage of security protection strategies (e.g., ARP prevention, AP isolation, and Rogue AP detection), an off-path attacker would not be able to discern if any client is communicating with a specific remote server. Furthermore, the attacker would not be able to ascertain the source port of the TCP connection, if it exists, and the sequence and acknowledgment numbers.}
\NewMajor{
%
%Consequently, our attack is particularly meaningful against enterprise Wi-Fi networks, which sets our attack apart from the rogue clone attacks, although our attack enables a new approach to compromising open and home mode networks enabled with defense measures against the existing attacks.
}

% To successfully launch \Major{our novel} \xw{(Why we emphasize ``our novel''? remove it?)} attacks, there are some requirements to be fulfilled. 
To successfully launch our attacks, there are some requirements to be fulfilled. 
% First, the attacker should be able to probe potential victim clients and the external IP address of the router. In IPv4 networks, the attacker can easily finish this task with many network scanning tools (e.g., Nmap \cite{nmap}, Traceroute \cite{traceroute}, Ping \cite{ping}). Previous research has revealed that nearly 89\% of the public Wi-Fi networks disable AP isolation and allow clients in the same LAN to communicate with each other \cite{feng23mitm}. 
First, the attacker should be able to probe the external IP address of the router. \Major{We will illustrate our methods in Section \ref{sec:network_probe}. 
Then the attacker tests whether AP isolation is enabled in the network\footnote{\xw{Previous research has revealed that nearly 89\% of the public Wi-Fi networks allow clients to communicate with each other \cite{feng23mitm}.}}~\cite{AP-isolation-detection}.
The attacker can successfully carry out the TCP DoS and TCP hijacking attacks regardless of whether AP isolation is enabled.
\xw{When AP isolation is enabled, the attacker will not be able to probe potential victim clients within the network using scanning tools (e.g., Nmap \cite{nmap}). Thus, the TCP injection attack will be thwarted with AP isolation. We will discuss the impacts of AP isolation on our attacks in Section \ref{sec:hijack} and Section \ref{sec:practical_consideration}.}}
Besides, the attacker has to target the remote server that the client is communicating with or will connect to, which can be set as those providing popular services as previous works \cite{cao2016off, ccsfeng}, e.g., famous servers, web search engines, or social sites. 
%The advantage in Wi-Fi networks is that the attacker and victims are in geographical proximity then it is possible for the attacker to get the potential servers with social engineering methods.

Second, the router adopts the \textit{port preservation} strategy.
% when creating new NAT mappings of TCP protocol and will assign another source port once collision happens. 
Our investigations show that most routers adopt this strategy except for an enterprise router model from Huawei \Major{and the open-sourced routing firmware of pfSense~\cite{pfSense}}. 
%
%question: whether to mention huawei AR6140E-9G-2AC？
%
Also, the router disables the reverse path validation strategy. 
We find that routers from 24 out of 30 vendors will forward forged packets except for Asus, Aruba, Cisco Meraki, Netgear, pfSense, and ZTE. Besides, some models from TP-Link, Mercury and Huawei also enable this strategy. 
% and thus are not influenced by the attack.
%
Moreover, the router disables the TCP window tracking strategy.
% which allows an attacker to clean the NAT mappings of other clients with forged TCP \texttt{RST} packets and construct a new NAT mapping at the router with a forged TCP data packet. 
In our measurement, most routers have disabled it by default, with the exception of Cisco Meraki.

Third, the victim client does not communicate with the server frequently. The state of the NAT mapping will transfer from \texttt{ESTABLISHED} to \texttt{CLOSE} state after receiving corresponding TCP \texttt{RST} packets, and the mapping will be removed completely after its timeout (1 second or 10 seconds in our test). It should be noted that if the client's communication continues during this period, it may interfere with the attack as the mapping will be refreshed.
%Additionally, if the client communicates with the server after the attacker constructs a new mapping occupying the source port, the connection will be \Major{terminated} from the client side, which causes a failure when the attacker has intended to inject fake data into the connection. Although it can be seen as a success when the attacker has intended to launch a DoS attack.
As there are many long-lived TCP connections that clients periodically retrieve new data from the server in minutes and 42\% of the tested routers set the timeout to only 1 second, the attacker has been provided with enough time to finish its attack. We will analyze its influence in detail in Section \ref{sec:practical_consideration}.

% \item {The attacker can send spoofed packets using the AP as the source IP address. IP spoofing is possible when the attacker and the victim originator reside in the same network. Because the security features that block spoofed packets are often deployed at gateways (higher layers of aggregation) to filter the network traffic flowing through\cite{rfc2827}, the spoofed TCP packets internally do not pass through the gateway and thus are not subject to blocking. Besides, according to our studies on \55 popular AP routers, we find that none of them can prevent the attacker to send spoofed packets in the same LAN, even when the spoofed messages use the remote server as the source IP address.}

%-------------------------------------------------------------------------------
\section{Attack Procedure}\label{sec:attack_procedure}
\subsection{Attack Overview}\label{sec:attack_overview}

% The victim client and the attacker are adjacent users who reside in the same Wi-Fi network in public places (e.g., coffee shops, restaurants, hotels, and shopping malls). The victim client may enjoy the Internet service, i.e., visiting web pages, \Major{downloading files}, or using the SSH service to control remote hosts. 
% With the wide deployment of modern security protection mechanisms such as WPAs, the traffic originating from the victim client is encrypted by a session key between the router and the victim. The attacker may sniff the encrypted frames, but it cannot decrypt the frames without knowing the particular session key. In this way, the off-path attacker cannot know whether any client is communicating with a given remote host, or the source port that the client is using associated with any existing TCP connection, let alone the sequence and acknowledgment numbers of the TCP connection.

% However, in this paper, we reveal that an off-path network-adjacent attacker can infer the existence of connections from the LAN to a given remote host, obtain the sequence and acknowledgment numbers of the existing TCP connection, and reset the connection with a TCP \texttt{RST} packet or inject arbitrary data into the connection if there is no additional encryption at the application layer (e.g., SSL/TLS). Besides, it may hijack the traffic directly by rerouting the responses from the server to itself and thus get confidential information of the victim.

To perform our attacks, the attacker has to carry out the following three steps:

\begin{itemize}
% [leftmargin=*]
\item[1.] Probe the router's external IP address and identify \Major{whether AP isolation is enabled, thus finding potential victim clients}.
\item[2.] Make inferences about whether there is any active connection from the LAN to the server.

\item[3.] Evict and construct NAT mappings at the router and then intercept the sequence and acknowledgment numbers from the replies to unsolicited packets from the server. 

\end{itemize}

\Major{
After the above steps, the attacker can terminate the connection directly or hijack the connection by replacing the victim client.
Besides, when AP isolation is disabled, the attacker can restore the original NAT mapping of the victim client at the router and send fake response packets to the client.}

\subsection{Phase 1: Probing the Network}\label{sec:network_probe}
In this step, the attacker prepares the attack in two aspects, namely, identifying the status of AP isolation in the network and probing the external IP address of the router. 
% Firstly, the attacker detects live hosts in the same network by leveraging some famous network scanning tools (e.g., Nmap \cite{nmap}, Ping \cite{ping}). 
Firstly, the attacker detects whether AP isolation is enabled via network scanning tools (e.g., Nmap \cite{nmap}, MacStealer \cite{AP-isolation-detection}). 
If it is disabled, the attacker records the scanning results of potential victim clients for the futural TCP injection attack.
Note that the attacker does not need to know the specific private IP address of the victim client (i.e., which IP is the victim), as we will show that it only needs to send related packets to the router, the server, or all of the clients in the subsequent attack phases, which is different from previous works that they will choose a target victim client beforehand, i.e., identifying whether a given client is communicating with the server \cite{cao2016off,ccsfeng} or redirecting the victim's traffic to the attacker \cite{feng23mitm}.
% If the attacker needs to inject TCP packets to the victim client, it can send the packets to the external IP of the router by abusing the disabled reverse path validation and the NAT mappings on the router (as the same as we did when inferring the source port).
 
% Our empirical study finds that we can detect live hosts in 90\% of the public Wi-Fi networks as they do not enable AP isolation.

% Secondly, the attacker probes the router's information, such as its brand, OS type, and external IP address. The attacker can access the gateway's IP address in the web browser and in most cases it will be the router's setting page (Web GUI) which indicates the router's brand. And the attacker can also leverage the Nmap tool to infer the router's OS type with its provided option. With the brand and OS type of the router, the attacker can judge in advance whether the router is affected by the attack, and determine the relevant parameters in subsequent attack steps through previous laboratory test results, i.e., sequence numbers to be specified in the TCP \texttt{RST} packets and expiration timeouts of NAT mappings. 
Secondly, the attacker probes the router's external IP address. With the widely deployed carrier-grade NAT \cite{imc16_cgn}, the Wi-Fi networks in the real world may consist of multiple levels of NAT \cite{RFC5684}, which means that the router's external IP address is not always a public IP address that can be obtained easily by querying its own public IP. 
\Major{We adopt the following methods to deal with this problem.} 
(i) First, the attacker gets the gateways along the way to any outside host (e.g., 8.8.8.8) through Traceroute \cite{traceroute}. 
Second, the attacker issues the ping command to the second gateway with the \textit{RECORD\_ROUTE} option, which will record the passed routes \cite{ping}, and then all the IP addresses of the passed interfaces will be returned. 
\Major{\xw{The result snapshot of the method is provided in Appendix \ref{appendix:probe-IP} (refer to Figure~\ref{fig:probeExternalIP}).
(ii) In certain scenarios, the aforementioned method may encounter failure, as the passed routes might not be returned when pinging the second gateway. In such cases, the attacker can opt to scan the subnet of the second gateway to identify live hosts' IP addresses. Subsequently, it can proceed to ping these IPs using the \textit{RECORD\_ROUTE} option. When the ping reaches the external IP of the router, the previously passed routes will be returned. However, when pinging other IPs, the routes will not be returned.
Besides, the attacker can access these IPs via a web browser. When accessing the external IP of the router, the router's setting page (Web GUI) will be displayed, whereas accessing the other IPs will lead to different pages.}}

%\Major{The result's snapshot for this method is shown in Appendix \ref{appendix:probe-IP} (see Figure \ref{externalIP}). 
%
%(ii) In some cases, the method above may fail as the passed routes will not be returned when pinging the second gateway. The attacker can scan the subnet of the second gateway to get live hosts' IPs. Then it can continue to ping these IPs with the \textit{RECORD\_ROUTE} option. When it happens to ping the external IP of the router, the routes passed will be returned. While when pinging others, the routes will not be returned. 
%Besides, the attacker can access these IPs in the web browser. When accessing the external IP of the router, there will be the router's setting page (Web GUI) while pages will be different when accessing the others.
%}
% In this way, the attacker can obtain the external IP address of the connected router. 
% However, the attacker can deal with this problem in the following methods. First, the attacker can get the gateways along the way to any outside host (e.g., 8.8.8.8) through Traceroute \cite{traceroute}. 
% Note that only the first two gateways are useful. 
% Second, the attacker sends ICMP echo request packets to the second gateway with the \textit{RECORD\_ROUTE} option which will record the passed routes \cite{ping}, and then all the IP addresses of the passed interfaces will be returned. In this way, the attacker can obtain the external IP address of the connected router.
% , which will be used for the following attacks.

\subsection{Phase 2: Making Inferences about Active Connections}\label{sec:connection}

Assuming that the attacker has connected to a Wi-Fi network, in which one of the normal users has established a TCP connection with a remote server from source port \texttt{m}. The router has a corresponding NAT mapping to keep track of the connection.
% and record the related information of the connection. The attacker intends to infer whether there is any active connection from the LAN to a given outside server and which source port is used by the victim client if there is. 
%
The attacker intends to infer which source port is used by the victim client of the connection.

\begin{figure}[ht]
	% \vspace{-1mm}
	\begin{center}
            \vspace{7pt}	%修改为7pt
		\includegraphics[width=\linewidth]{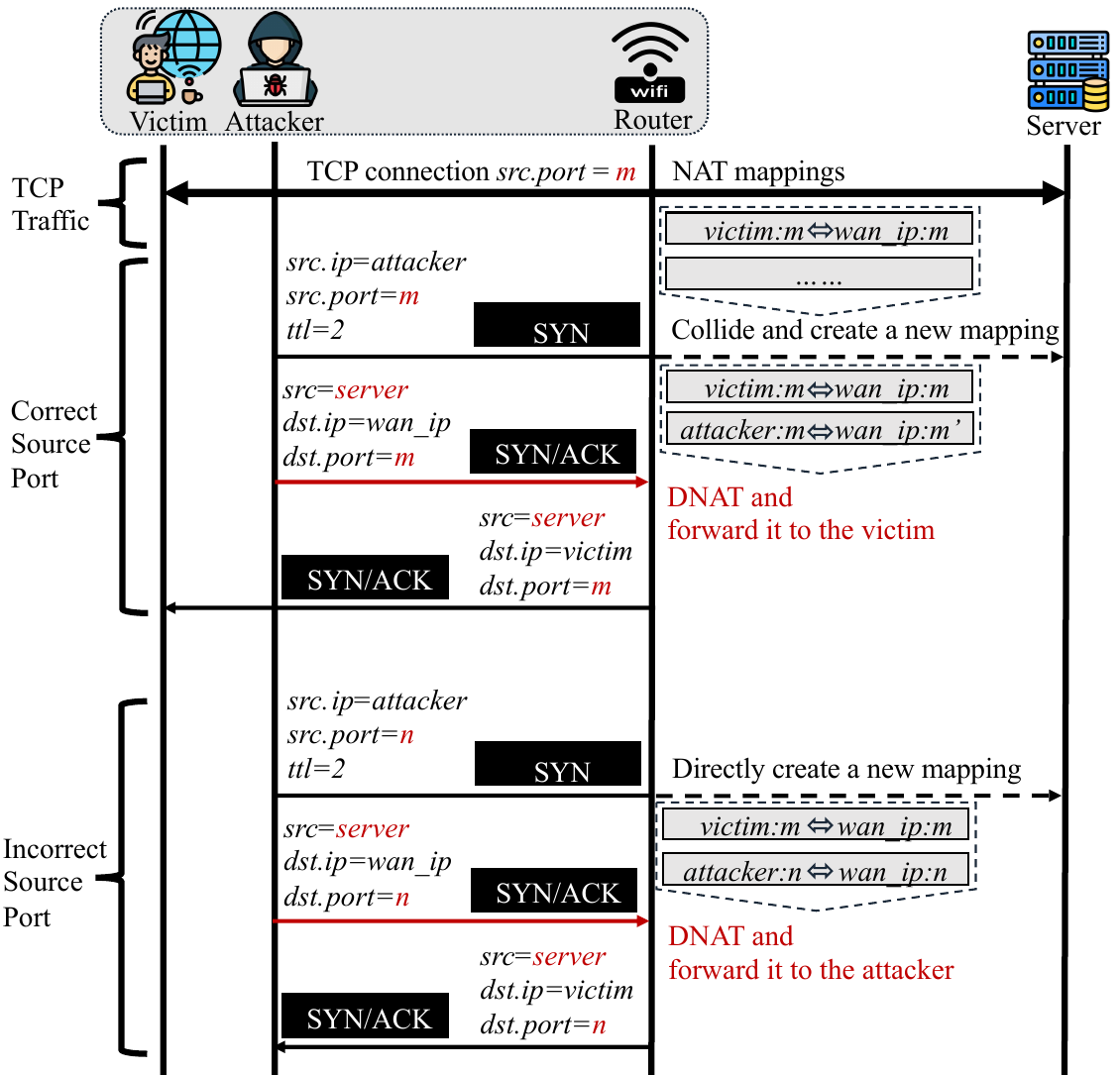}
		% \vspace{2mm}
		\caption{Inferring the source port of the victim TCP connection}
		\label{connectioninference}
	\end{center}
	% \vspace{-4mm}
\end{figure}

Figure~\ref{connectioninference} illustrates the side-channel vulnerability that leverages the NAT \textit{port preservation} strategy and insufficient reverse path validation of the router. First, the attacker sends a \texttt{SYN} packet targeting the server with its own IP address and a guessed port number as the source. If the source port number (e.g., \texttt{n}) specified in the \texttt{SYN} packet does not equal \texttt{m}, the router will create a new NAT mapping with source port \texttt{n} that records this new TCP connection.
% and translates the subsequent packets between the attacker's private IP address and the router's external IP address. 
%
Note that with the wide deployment of NIDS at the server side \cite{NIDS}, a large amount of \texttt{SYN} packets arriving at the server may be detected, and the server may find it attacked and take corresponding fightback. The attacker can set the \texttt{TTL} of the \texttt{SYN} packet to a small number (e.g., 2 in our test), and thus the packet will be dropped quickly at the intermediate routers. In most cases, the routers do not deploy detection systems typically. 

Then, the attacker impersonates the server and sends a forged \texttt{SYN/ACK} packet whose destination is the router's external IP address and whose destination port is the guessed source port \texttt{n}. According to RFC 3704 recommendation, the reverse path of packets received should be strictly checked so as to prevent IP spoofing attacks. In this case, as the \texttt{SYN/ACK} packet is received from the router's internal interface while its source address is a public IP address and it actually cannot be routed from this incoming interface for responding to the packet, then it should be dropped by the router. However, many routers in the real world do not adopt the RFC recommendation and will not check the reverse path of packets received. So when the forged packet arrives, it will be processed by the router's kernel and forwarded according to NAT mappings. Since there is a NAT mapping that translates the router's external IP to the attacker's private IP when the source is the remote server and the destination port is \texttt{n}, the forged packet will match this mapping and be forwarded to the attacker. In this way, the attacker can receive the forged packet that is sent from itself again if the guessed source port \texttt{n} is not equal to the victim client's source port \texttt{m}.

If the attacker guesses the right source port, i.e., \texttt{m}, when the \texttt{SYN} packet arrives at the router, it will translate the source port of the new mapping to another port due to the collision. Let us say that the changed source port is \texttt{m'}. In the second step, when the forged \texttt{SYN/ACK} arrives at the router, however, it will be forwarded to the victim according to the client's NAT mapping as the port specified in it is \texttt{m} instead of \texttt{m'}. Thus, from the view of the attacker, it cannot receive the forged \texttt{SYN/ACK} packet again if the port it guesses is right, i.e., previously occupied. In this way, the attacker can infer that there is a connection from some local host to the target server with the source port \texttt{m}. 

The attacker repeats the above procedure, i.e., changing the
guessed source port number specified in the forged \texttt{SYN} and \texttt{SYN/ACK} packets and then observing if it can receive \texttt{SYN/ACK} back until the correct port \texttt{m} is identified, which will be used for the subsequent attacks.

\subsection{Phase 3: Hijacking Active Connections}\label{sec:hijack}
Once the attacker has determined an active TCP connection to a given remote server with the source port \texttt{m}, it will attempt to obtain the current sequence number \textit{SEQ} and the acknowledgment number \textit{ACK} of the server. Note that as TCP is a bidirectional symmetric full-duplex protocol, the sequence and acknowledgment numbers of the victim client are symmetrical to the server.
% Previous works mostly infer these values by exploring the entire possible space, i.e., sending spoofed packets with guessed numbers to the server or the client and observing unintended changes of some fields (e.g., shared IPID counter \cite{ccsfeng}, global value of challenge ACK limit \cite{cao2016off}, the Wi-Fi timing difference \cite{chen2018off}, or the size of triggered packets in VPN scenarios \cite{Tolley21VPN}). As sequence and acknowledgment numbers are 32-bit random numbers, it will take a lot of effort to traverse the entire 4G space, which largely impacts the success rate of hijacking the short-period TCP connections. 
Previous works mostly infer these values by exploring the entire possible 4G space via leveraging some side channels \cite{ccsfeng,cao2016off,chen2018off, Tolley21VPN}), which are rather time-consuming and largely impact the success rate of hijacking the short-period TCP connections. 
However, in this work, we demonstrate a new method to obtain these two values directly and precisely, which abuses vulnerable routers without TCP window tracking.
% Assuming that there's a TCP connection from a local client to a remote server with the source port of \texttt{m}, 
We assume that there will be some intervals in the communication between the client and the server. Depending on the scenarios, the client periodically initiates a request and waits for responses, or the server proactively pushes notification messages, which are often the cases in real-world services.

\begin{figure}[h]
	% \vspace{-2mm}
	\begin{center}
            \vspace{7pt}	%修改为7pt
		\includegraphics[width=\linewidth]{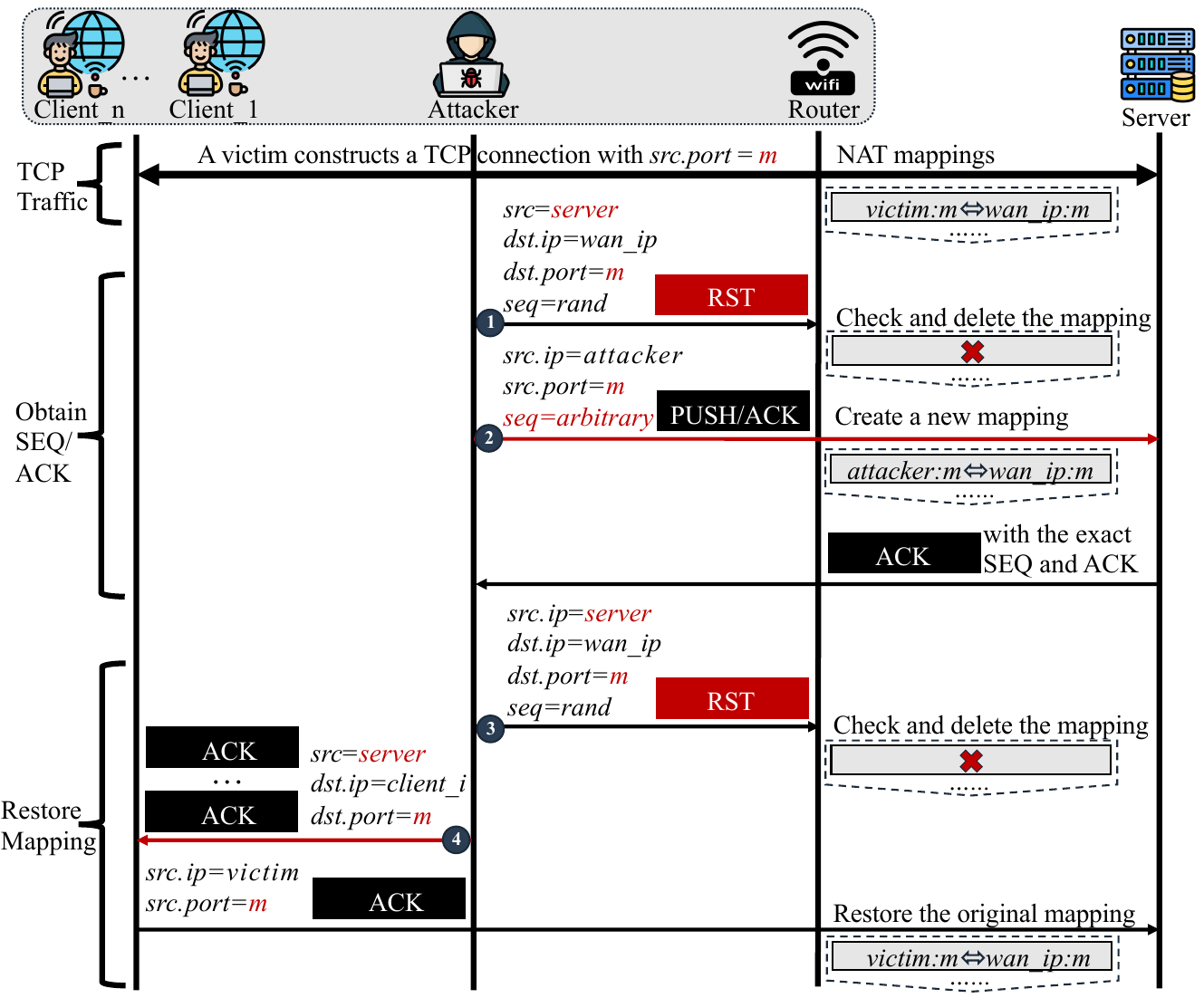}
		% \vspace{2mm}
		\caption{Hijacking active connections}
		\label{resetmapping}
	\end{center}
	% \vspace{-4mm}
\end{figure}

Figure~\ref{resetmapping} shows our method for the attacker to hijack the TCP connection between the victim client and server. 
% The method contains four steps. 
Firstly, the attacker cleans the router's NAT mapping of the victim connection by sending spoofed TCP \texttt{RST} packets whose source is the server and destination IP is the router's external IP, and destination port is the previously inferred port \texttt{m}. The sequence numbers specified in these packets are crafted for various brands and types of routers due to their different behaviors of disabling the TCP window tracking strategy. Generally, there are two popular behaviors as stated in Section \ref{sec:tcp_window_tracking} that the first type of router does not check the sequence number at all, and the second type of router will check if the sequence number is beyond the exact sequence number in a 2G space. 
% More of the specific information on the vulnerable routers will be detailed in \ref{sec:Vulnerable_Routers}. 
For the first type of router, the attacker can specify an arbitrary sequence number in one crafted \texttt{RST} packet to clean the TCP NAT mapping at the router directly. And for the second type of router the attacker can send two \texttt{RST} packets, one with sequence number \textit{x} and the other with \textit{(x + 2G) \% 4G}, which ensures that one of them will fall within the required range.

To simplify the description, we will only elaborate on the details of the attack on the first type of router. After receiving the \texttt{RST} packet, the router will falsely process the packet as it does not perform reverse path validation, and the state of the recorded TCP mapping will transfer from \texttt{ESTABLISHED} to \texttt{CLOSE}. The \texttt{CLOSE} state will last for a timeout according to a kernel variable (e.g., \textit{ip\_conntrack\_tcp\_timeout\_close} in OpenWrt-based systems). In our empirical investigation, 41\% of the tested routers set this value to as short as 1 second, and the other vulnerable routers set it to 10 seconds. After the countdown of the timeout, the original TCP NAT mapping at the router will be completely cleared.

Secondly, after the original NAT mapping has been fully evicted, the attacker replaces the victim client by constructing a new mapping at the router via sending a forged data packet with \texttt{PUSH/ACK} flags to the remote server, using the port \texttt{m} as the source port and its private IP as the source IP address. The sequence and acknowledgment numbers specified in the packet can be arbitrary as the router does not verify them. 
With the translation of IP addresses at the router, the packet will be routed to the server. From the perspective of the remote server, this packet is from the public IP address of the client with source port \texttt{m} and it will match the victim connection. Since the sequence and acknowledgment numbers specified in the packet are wrong, an \texttt{ACK} packet with the server's exact sequence and acknowledgment numbers will be triggered back \cite{rfc793}. When the \texttt{ACK} packet arrives at the router, it will be translated and routed to the attacker according to the new NAT mapping created just now. Then it can obtain the sequence and acknowledgment numbers of the victim connection directly.

After the above two steps, the attacker can decide on the follow-up procedures according to the purpose of the attack. 
\Major{In this work, we will illustrate three types of possible attacks. \textbf{(i) TCP DoS attack.} If the attacker intends to forcibly close the connection in the scenario of encrypted tunnels (e.g., SSH or HTTPS), it can just send forged TCP \texttt{RST} packets to the server with the information obtained before, thus causing the connection terminated at the server side. After that, the client will not receive any response when it sends requests to the server, which leads to a denial of service attack. 
\textbf{(ii) TCP hijacking attack.} If the attacker intends to hijack the traffic from the server to the victim client, it can take over the NAT mapping and impersonate the client again with the exact sequence and acknowledgment numbers to launch requests to the server and wait for responses from the server.
% If the servers only authenticate users during the initial login phase, this attack may lead to permission bypass and privacy leakage. 
% We will detail this case in ~\ref{sec:mitm_discussion}.
For instance, at the beginning, the victim client logins into the FTP server and requests personal files from the server. After the attack, the attacker can bypass the initial verification stage by replacing the victim client to send requests to the server, which may lead to permission bypass and privacy leakage.
% We will deeply dive into this kind of attack and how to make it well applicable to real-world applications is what we may consider in the future.
\textbf{(iii) TCP injection attack.} 
If the attacker intends to send forged responses by impersonating the server when the victim client initiates a new request later, it needs to restore the original NAT mapping of the victim client at the router so as not to interfere with the client's normal communication. We are going to elaborate on this case as shown in the last two steps in Figure~\ref{resetmapping}.
}

% The attacker has to clean its NAT mapping and reconstruct the mapping of the victim. It
The attacker repeats the first step to clean the mapping of itself and waits for another NAT mapping timeout of \texttt{CLOSE} state.
To restore the original mapping of the victim, as the attacker does not know the victim client's exact private IP, it can send forged \texttt{ACK} packets to all of the local hosts probed in the network probing phase (see Section \ref{sec:network_probe}) when AP isolation is disabled. 
The source of the \texttt{ACK} packets is the server, and the destination port is \texttt{m}. After arriving at the irrelevant hosts with no corresponding connections, the packets will be dropped. Yet the victim will send an \texttt{ACK} packet back to the server, which restores the NAT mapping of the victim when it travels through the router. 
Then the attacker can inject forged responses to the client by sending them to the external IP of the router via abusing the disabled reverse path validation and NAT mappings (the same as it did when inferring the source port).
However, the attacker cannot restore the mapping if AP isolation is enabled within the network, as it cannot send packets to other clients, and thus this attack is thwarted.

Note that in the \Major{above} attack phases, there are timeouts in which the NAT mappings are in the state of \texttt{CLOSE}. We have found that if related packets are traveling through the router at this period, the countdown will be refreshed, which may lead to interference with the attack. We will discuss the influence in detail in Section \ref{sec:practical_consideration}.

%-------------------------------------------------------------------------------
\section{Empirical Study}\label{sec:empirical_study}

We conduct extensive real-world evaluations to measure the impacts of the attack. We first investigate the default settings of routers on the market.
% , including their port allocation strategies, whether they enable reverse path validation and TCP window tracking, and timeouts of TCP \texttt{CLOSE} state. 
\Major{Next, we conduct case studies to evaluate the effectiveness of the attack in various real-world Wi-Fi networks.}
% We have tested 59 popular router models from 22 vendors and found that 48 of them from 19 vendors are vulnerable to the attack. Next, we conduct two case studies to evaluate the effectiveness and efficiency of the attack in 93 real-world Wi-Fi networks and discover that 75 of them (81\%) may suffer from the attack. 
% \Major{For the TCP DoS attack, an off-path attacker can reset a TCP connection in 18.1 seconds and the success rate is 87.4\% on average. For the TCP hijacking attack, it can request a private file from the FTP server in xxx seconds and the success rate is xxx\% on average. And for the TCP injection attack, it can manipulate the web traffic within 54.5 seconds and the success rate is 76.1\% on average.}

%\subsection{Ethical Considerations}\label{sec:ethical_considerations}
\noindent\textbf{Ethical Considerations.}
As it is essential to respect the privacy and security of others when engaging in authorized hacking experiments, our experiments in real-world Wi-Fi networks require careful consideration of ethical issues.
%Prior to our experiment, we obtained the consent of the Wi-Fi network administrator and carried out the experiments in the absence of any users. We also make sure that all testing devices were our own controlled devices to avoid affecting possible uncontrolled user devices. After our experiment, we promptly notified the affected Wi-Fi network administrator of relevant vulnerabilities and proposed mitigation measures. And once the experiment finishes, the network administrator reboots the router to clear irrelevant caches generated during the experiment to prevent affecting subsequent usage.
%
We addressed the ethical issues of our real-world experiments from the following perspectives. 
First, we provided the Wi-Fi network administrators with detailed explanations of our experimental plans and obtained their approval before conducting the experiments. 
Second, with the help of the administrator, we ensured that no other users were accessing the Wi-Fi network during our experiments, thus avoiding potential risks or side effects for other users. 
We then deployed our machine (a laptop or a cellphone) as the victim client in the Wi-Fi network, thus ensuring that all the machines involved in our experiments were under our control and would not affect other machines.
Finally, after completing the experiments, we provided feedback on the results to the administrator. Moreover, we recommended that they restart the Wi-Fi router and clear the cache to restore the network to a safe state.

\subsection{Analysis of Routers}\label{sec:Vulnerable_Routers}

% The attack leverages the strategies of the router adopted and only if all of the conditions are fulfilled can our attack succeed. In order to explore the coverage of vulnerable routers, we perform tests on real router models from leading vendors including TP-Link, D-Link, Tenda, ASUS, Netgear, Huawei and Linksys, and other prominent players like Mercury, Xiaomi, Ruijie, ZTE and H3C, as well as Wavlink, Comfast, IP-COM, Skyworth, Volans, Netcore, 360, iKuai, WiMaster, China Mobile. In total, we perform tests on 59 mainstream router models (acting as gateway to provide Internet access services) from 22 vendors. For each router model, we test if it fits all attack conditions proposed in \ref{sec:threat_model}.
%
The attack leverages the strategies adopted by the router, and only if all of the conditions are fulfilled can our attack succeed. In order to explore the coverage of vulnerable routers, \Major{we perform tests on real router models from lots of vendors, including 
360, Aruba, ASUS, Amazon, Cisco Meraki, China Mobile, Comfast, D-Link, GL.iNet, Google, H3C, Huawei, IP-COM, iKuai, JdCloud, Linksys, Mercury, Netgear, Netcore, Ruijie, Skyworth, Tenda, TP-Link, Ubiquiti, Volans, Wavlink, WiMaster, Xiaomi, and ZTE.
% TP-Link, D-Link, Ubiquiti, Aruba, Cisco Meraki, Amazon, Google, Tenda, ASUS, Netgear, Huawei and Linksys, Mercury, Xiaomi, Ruijie, ZTE and H3C, as well as Wavlink, Comfast, IP-COM, Skyworth, Volans, Netcore, 360, iKuai, WiMaster, GL.iNet, JdCloud, China Mobile.
To our best knowledge, the operating systems of most routers we tested are based on Linux with custom modifications, except for some router models from TP-Link and Mercury, which are based on VxWorks. Therefore, we also build a soft routing environment with a FreeBSD-based firmware, i.e., pfSense 2.7.0~\cite{pfSense}.} In total, we perform tests on \Major{67} mainstream router models (acting as the gateway to provide Internet services) from 30 vendors. For each router model, we test if it fits all attack conditions proposed in Section \ref{sec:threat_model}. Here we list the detailed test results of 33 representative routers from these 30 vendors in Table~\ref{part-routers}.

First, the router has to take the \textit{port preservation} strategy.
% when allocating new TCP NAT mappings, and will choose another TCP port with particular methods when a collision happens.
In our test, most of the routers adopt this strategy by default. Only the enterprise wired router model ``AR6140E-9G-2AC'' produced by Huawei \Major{and the soft routing machine with pfSense} which co-work with a wireless AP to provide Wi-Fi service, take the \textit{random selection} strategy that prevents the attacker from inferring the source port of clients' connections. Besides, there is no router model which takes the \textit{sequential selection} or \textit{port overloading} strategy as stated in Section \ref{sec:background-NAT}

%As the kernels of most routers are based of Linux, such as Asuswrt\cite{} and Openwrt\cite{}, we find that in the implementation of Linux Netfilter module\cite{} will choose the port....

Second, we investigate the deployment of the reverse path validation strategy in the routers. Among these routers, Netgear and Asus set the kernel variable \textit{rp\_filter} to 1 by default, which means they are secure to check the packet received strictly. And some of the old-styled models of TP-Link and Mercury \Major{(i.e., VxWorks-based)} will also validate the received packets. However, their newest models (e.g., designed for Wi-Fi 6) and some enterprise routers will not validate them anymore. In addition, ZTE routers, the router model ``AR6140E-9G-2AC'' from Huawei, \Major{and the soft routing machine with pfSense} will also not forward the forged packets. The routers from the other vendors all disable reverse path validation, which results in the vulnerability of inferring active connections.

Third, the router has to disable the TCP window tracking strategy.
% so that forged \texttt{RST} packets can be abused to evict the NAT mappings.
% After a systematic investigation of two famous open-source router operating systems (Asuswrt and Openwrt), we found that they both have related options to set or disable the TCP window tracking strategy. 
In our test, \Major{only one enterprise wired router model ``Meraki 64'' produced by Cisco Meraki will check the sequence number strictly and} all of the other routers disable TCP window tracking while the processing logic is slightly different. Asus, Google, Netgear, Tenda, Wimaster, and ZTE routers will check if the sequence number is in the 2G space beyond the exact sequence number.
% , which is also very loose and can be bypassed by sending two \texttt{RST} packets whose sequence numbers have a gap of 2G. 
And the other routers do not check the sequence number at all.
% thus a \texttt{RST} packet with an arbitrary sequence number can evict the NAT mapping at the router.

 % When an attacker sends a forged TCP reset packet to clean the NAT mappings, the Openwrt-based routers will not check the sequence number specified within it while the Asuswrt-based routers will check if it is in the 2G space beyond the current sequence number,  As for some VxWorks-based routers (e.g., some models of TP-Link and Mercury), their codes are not open source and we don't know the specific implementation of this strategy. But in our test, their behavior is similar to the Openwrt-based models that they will not check the sequence number too.

Fourth, the timeout of TCP \texttt{CLOSE} state of NAT mappings may influence the time cost and success rate of our attack.
% , which is the time for mappings to be completely cleaned after receiving \texttt{RST} packets. 
The shorter the timeout is, the easier the attack can succeed. 
% In our test, we find that the default timeout of the old-style routers is 1 second and the newest models of these brands choose to update this time to 10 seconds\cite{netfilter}. 
Among the 66 router models without TCP window tracking, 28 of them will clean the NAT mapping in only 1 second, 37 of them will be tricked to clean the mapping in 10 seconds, and the default setting of pfSense is 90 seconds. 
%We speculate that these devices are in a higher version of Netfilter \cite{netfilter} as its current default setting is 10 seconds.

\begin{table*}[t]
\small
\renewcommand\arraystretch{1.1}
\setlength\tabcolsep{1pt}
\centering
\caption{Partial Tested Routers From 30 Vendors}
\label{part-routers}
%\resizebox{\linewidth}{!}{%
\begin{threeparttable}

\scalebox{0.8}{
\setlength{\tabcolsep}{2mm}{
\begin{tabular}{l|l|ccc|cccc|c} 
\bottomrule

\textbf{No.} & \textbf{Router Model} & \textbf{Vendor } & \textbf{OS} & \textbf{Generation }   
& \textbf{\tabincell{c}{Port \\Preservation}}
& \textbf{\tabincell{c}{Reverse-path \\Validation \\ Disabled}}
& \textbf{\tabincell{c}{TCP Window \\Tracking \\ Disabled}}
& \textbf{\tabincell{c}{TCP Close \\Timeout \\(second)}}
& \textbf{Vulnerable} \\ 

\hline

1 & TL-XDR6020 & TP-Link & Linux-based & Wi-Fi 6  & \blackcheck & \blackcheck & \blackcheck  &1 & \redcheck \\
2 & TL-WDR7620 & TP-Link & Vxworks-based & Wi-Fi 5  & \blackcheck & \blackcross & \blackcheck &1  & \greencross \\
\hline

\rowcolor[rgb]{0.871,0.871,0.871} 3 & AX3 Pro & Huawei & EMUI (Linux-based) & Wi-Fi 6 & \blackcheck & \blackcheck & \blackcheck & 10 & \redcheck \\
\rowcolor[rgb]{0.871,0.871,0.871} 4 & AR6140E-9G-2AC$^*$ & Huawei & VRP (Linux-based) & -    & \blackcross & \blackcross & \blackcheck &10 & \greencross \\
\hline

5 & V6G  & 360 & 360OS(Linux-based) & Wi-Fi 6  & \blackcheck & \blackcheck & \blackcheck &1 & \redcheck \\
\hline

\rowcolor[rgb]{0.871,0.871,0.871} 6 & Magic R365  & H3C & Comware(Linux-based) & Wi-Fi 5  & \blackcheck & \blackcheck & \blackcheck &10 & \redcheck \\
\hline

7 & W30E  & Tenda & Linux-based & Wi-Fi 6 & \blackcheck & \blackcheck & \blackcheck & 1 & \redcheck \\
\hline

\rowcolor[rgb]{0.871,0.871,0.871} 8 & RAX1800Z  & China Mobile & AOS(Linux-based) & Wi-Fi 6  & \blackcheck & \blackcheck & \blackcheck & 10 & \redcheck \\
\hline

9 & X32 Pro & Ruijie & RGOS(Linux-based) & Wi-Fi 6  & \blackcheck & \blackcheck & \blackcheck  &1 & \redcheck \\
\hline
% openwrt
\rowcolor[rgb]{0.871,0.871,0.871} 10 & Redmi RA81  & Xiaomi & MiWiFi(Linux-based) & Wi-Fi 6  & \blackcheck & \blackcheck & \blackcheck &1 & \redcheck \\
\hline

11 & MW300R & Mercury & Vxworks-based & Wi-Fi 4   & \blackcheck & \blackcross & \blackcheck &1 & \greencross \\
12 & X30G & Mercury & Linux-based & Wi-Fi 6  & \blackcheck & \blackcheck & \blackcheck &1 & \redcheck \\
\hline

\rowcolor[rgb]{0.871,0.871,0.871} 13 & RAX50  & Netgear & DumaOS(Linux-based) & Wi-Fi 6  & \blackcheck & \blackcross & \blackcheck &10 & \greencross \\
\hline

14 & RT-AX89X & ASUS & AsusWrt(Linux-based) & Wi-Fi 6   & \blackcheck & \blackcross & \blackcheck &10 & \greencross \\
\hline

\rowcolor[rgb]{0.871,0.871,0.871} 15 & E9450 & Linksys & Linux-based & Wi-Fi 6  & \blackcheck & \blackcheck & \blackcheck &10 & \redcheck \\
\hline

16 & QUANTUM D2G & Wavlink & Linux-based & Wi-Fi 5 & \blackcheck & \blackcheck & \blackcheck &10 & \redcheck \\
\hline

% \rowcolor[rgb]{0.871,0.871,0.871} 46 & CF-616AC & Comfast & OrangeOS(OpenWrt-based) & Wi-Fi 5  & \blackcheck & \blackcheck & \blackcheck &10 & \redcheck \\
% \hline
\rowcolor[rgb]{0.871,0.871,0.871} 17 & CF-616AC & Comfast & OrangeOS(Linux-based) & Wi-Fi 5  & \blackcheck & \blackcheck & \blackcheck &10 & \redcheck \\
\hline

18 & DI-7003GV2$^*$ & D-Link & Linux-based & -  & \blackcheck & \blackcheck & \blackcheck &1 & \redcheck \\
\hline

\rowcolor[rgb]{0.871,0.871,0.871} 19 & AX3000  & ZTE & ZXR10ROS(Linux-based) & Wi-Fi 6 & \blackcheck & \blackcross & \blackcheck &10 & \greencross \\
\hline

20 & M80$^*$ & IP-COM  & Linux-based & -  & \blackcheck & \blackcheck & \blackcheck &1 & \redcheck \\
\hline

\rowcolor[rgb]{0.871,0.871,0.871} 21 & SK-WR6640X  & Skyworth  & Linux-based & Wi-Fi 6   & \blackcheck & \blackcheck & \blackcheck &10 & \redcheck \\
\hline

22 & VE5200G$^*$ & Volans   & Linux-based & - & \blackcheck & \blackcheck & \blackcheck &1 & \redcheck \\
\hline

\rowcolor[rgb]{0.871,0.871,0.871} 23 & NBR1009GPE & Netcore  & NOS(Linux-based) & -  & \blackcheck & \blackcheck & \blackcheck &1 & \redcheck \\
\hline

% 56 & Wimaster-mini$^*$ & Wimaster  & OpenWrt-based & - & \blackcheck & \blackcheck & \blackcheck & 10 & \redcheck \\
% 57 & Wimaster$^*$ & Wimaster  & OpenWrt-based & - & \blackcheck & \blackcheck & \blackcheck & 10 & \redcheck \\
% \hline
24 & Wimaster$^*$ & Wimaster  & Linux-based & - & \blackcheck & \blackcheck & \blackcheck & 10 & \redcheck \\
\hline

% \rowcolor[rgb]{0.871,0.871,0.871} 58 & IK-Q90 & iKuai  & iKuaiOS(OpenWrt-based) & Wi-Fi 6  & \blackcheck & \blackcheck & \blackcheck &10 & \redcheck \\
% \rowcolor[rgb]{0.871,0.871,0.871} 59 & IK-Enterprise$^*$ & iKuai  & iKuaiOS(OpenWrt-based) & -  & \blackcheck & \blackcheck & \blackcheck &10 & \redcheck \\
% \hline
\rowcolor[rgb]{0.871,0.871,0.871} 25 & IK-Enterprise$^*$ & iKuai  & iKuaiOS(Linux-based) & -  & \blackcheck & \blackcheck & \blackcheck &10 & \redcheck \\
\hline

26 & Instant On AP22 & Aruba & ArubaOS(Linux-based) & Wi-Fi 6 & \blackcheck & \blackcross & \blackcheck & 10 & \greencross \\
\hline

\rowcolor[rgb]{0.871,0.871,0.871} 27 &  EdgeRouter X$^*$ & Ubiquiti  & Linux-based & - & \blackcheck & \blackcheck & \blackcheck &10 & \redcheck \\
\hline

28 & AX1800 & JdCloud  & Linux-based & Wi-Fi 6 & \blackcheck & \blackcheck & \blackcheck & 10 & \redcheck \\ 
\hline

\rowcolor[rgb]{0.871,0.871,0.871} 29 & Cisco Meraki 64$^*$ & Cisco Meraki & Linux-based & - & \blackcheck & \blackcross & \blackcross & - & \greencross \\
\hline

30 & eero pro & Amazon & Linux-based & Wi-Fi 5 & \blackcheck & \blackcheck & \blackcheck & 10 & \redcheck \\
\hline

\rowcolor[rgb]{0.871,0.871,0.871} 31 & Google Wi-Fi & Google  & ChromeOS(Linux-based) & Wi-Fi 5 & \blackcheck & \blackcheck & \blackcheck & 10 & \redcheck \\ 
\hline

32 & GL-MT3000 & GL.iNet  & Linux-based & Wi-Fi 6 & \blackcheck & \blackcheck & \blackcheck & 10 & \redcheck \\
\hline

\rowcolor[rgb]{0.871,0.871,0.871} 33 & pfSense 2.7.0$^*$ & pfSense & FreeBSD-based & - & \blackcross & \blackcross & \blackcheck & 90 & \greencross \\

\toprule
\end{tabular}}}
   \begin{tablenotes}
       \footnotesize
       \item[] \blackcheck means that the router is satisfied with the condition, and \blackcross means that the router is dissatisfied with the condition.
       \item[] \redcheck means that the router is vulnerable to our attack, and \greencross means that the router is immune to our attack.
       \item[] * means that the model is an enterprise router which does not support Wi-Fi by itself and needs to work together with wireless access points.
    \end{tablenotes}
\end{threeparttable}
%}
% \vspace{-5mm}
\end{table*}

%The detailed information of 30 out of the 59 tested routers is shown in Table~\ref{test-APs}.
Due to the limited space, the detailed information of the 67 tested routers is listed in Appendix~\ref{appendix:all_routers} (see Table~\ref{all-routers}).
We take the first row as an instance to analyze the results. The Linux-based router model ``TL-XDR6020'' produced by TP-Link, provides the latest generation of Wi-Fi 6 for network services. As for the four metrics mentioned above, this model takes the \textit{port preservation} strategy, does not validate the reverse path of received packets, disables the TCP window tracking, and sets the TCP \texttt{CLOSE} timeout to 10 seconds by default. In this way, it is vulnerable to our attacks. In conclusion, 52 of the 67 tested routers are vulnerable, and 15 models are immune to the attack as they do not fulfill all of the conditions. 
%, including the dominants in the router markets (e.g., TP-Link, Huawei, Linksys, and so on).
% And 15 models are immune to the novel attack.
%either because they obey the RFC recommendation and check the reverse path of packets strictly or because they take a \textit{random selection} strategy when allocating new mappings.

\subsection{Attack Evaluation}\label{sec:Attack_Evaluation}
% To evaluate the impacts of our attack in the real world, we also conduct thorough experiments of our attack in 93 various public Wi-Fi networks. We investigate whether the conditions of the attack are fulfilled in each network by taking two case studies of attacks on SSH and HTTP connections and measuring the time cost and success rate of each attack.

To evaluate the impacts of the attack in the real world, we also conduct thorough experiments of the attack in 93 various Wi-Fi networks. We investigate whether the conditions of the attack are fulfilled in each network by \Major{taking three case studies of attacks on SSH, FTP, and HTTP applications} and measuring the time cost and success rate of each attack.

\noindent \textbf{Experimental Setup}. Our experiments consist of four types of devices, i.e., a router, a victim client, a remote server, and an attacker.

\begin{itemize}
\item \textit{Router.} The router in Wi-Fi networks works as the gateway to provide Internet access and forward packets between local clients and outside servers. 

% \item \textit{Remote Server.} In the case of TCP DoS attacks, we set an SSH server equipped with Ubuntu 22.04 (kernel version 5.15.0), OpenSSH 8.9, and OpenSSL 3.0.2. In the case of TCP manipulation attacks, we pick a well-known finance website \textit{www.{ANONYMOUS}.com} (anonymized for ethical consideration) in which the client initiates a long-lived TCP connection that periodically retrieves data updates every minute. 
\Major{\item \textit{Remote Server.} For the DoS attack, we set up an SSH server equipped with Ubuntu 22.04 (kernel version 5.15.0), OpenSSH 8.9, and OpenSSL 3.0.2. For the hijacking attack, we set an FTP server equipped with Ubuntu 22.04 (kernel version 5.15.0) and vsftpd version 3.0.3. And for the injection attack, we pick a well-known finance website \textit{www.{ANONYMOUS}.com} (anonymized for ethical consideration) in which the client initiates a long-lived TCP connection that periodically retrieves data updates every minute.
}

% \item \textit{Victim client.} Though the OS type or version of the client is unrestricted in our attack, we still deploy victim clients equipped with five typical OSes (i.e., Windows, Linux, Mac, iOS, and Android). Each victim client is attached to the AP of the public Wi-Fi networks. In the case of TCP DoS attacks, the victim client will connect to the remote server through SSH. And in the case of TCP manipulation attacks, the victim client will access the website above to get the newest future index of stocks (e.g., \texttt{HSI},\texttt{HSCEI}).

\item \textit{Victim Client.} Though the OS type or version of the client is unrestricted in our attack, we still deploy victim clients equipped with five typical OSes (i.e., Windows, Linux, Mac, iOS, and Android). Each victim client has connected to the Wi-Fi network. \Major{In the case of DoS and hijacking attacks, the victim client will communicate with the remote server through SSH and FTP}. And in the case of the injection attack, the victim client will access the website above to get the newest future index of stocks (e.g., \texttt{HSI}, \texttt{HSCEI}).

\item \textit{Attacker.} An attack machine is equipped with Linux 5.15.0, which is capable of crafting packets. The attacker aims to \Major{forcibly close} the victim client's SSH connection with the remote server, \Major{steal private files from the FTP server,} or inject fake HTTP responses to the client by performing the attack. 

\end{itemize}
% \noindent \textbf{Attack Procedure}. The attacker first probes the Wi-Fi network to detect all the clients attached to the same \Major{network} and get the router's external IP address. Next, the attack can be constructed in the following steps: (1) detecting whether there is any TCP connection from the LAN to the given server, i.e., identifying the correct source port, (2) evicting the router's original NAT mapping of the client with forged \texttt{RST} packets and constructing a new one by sending a TCP data packet to the server, which in turn incurs an \texttt{ACK} packet from it, (3) breaking down the SSH connection directly, or restoring the original NAT mapping of the client. (4) answering the client's HTTP requests with forged segments specified with the inferred values.

\noindent \textbf{Attack Procedure}. The attacker first \Major{tries to} get the router's external IP address and test whether AP isolation is enabled and other hosts can be detected in the Wi-Fi network. Next, the attack can be constructed in the following steps: 
(1) detecting whether there is any TCP connection from the LAN to the given server, i.e., identifying the correct source port, 
(2) evicting the router's original NAT mapping of the client with forged \texttt{RST} packets and constructing a new one by sending a TCP data packet to the server, which in turn incurs an \texttt{ACK} packet from it, 
\Major{ 
(3) terminating the SSH connection or requesting an FTP file download,
(4) restoring the original NAT mapping of the client and answering the client's HTTP requests with forged segments specified with the inferred values.
}
%(3) and (4) should be at the same level

% We repeat the experiments 20 times in each tested Wi-Fi network. We define the result that the client or the server cannot receive the messages from each other as a successful \Major{\textbf{TCP DoS attack}}, which includes two cases. First, after the attacker receives the \texttt{ACK} packet from the server, the attacker can send a TCP \texttt{RST} packet with the exact sequence number to the server, resulting in the connection broken down at the server side. Second, as the attacker has replaced the NAT mapping at the router, if it happens that the client sends a packet to the server at this stage, the source port of the packet will be translated. When the packet arrives at the server, it will incur a reset packet from the server as there is no corresponding connection, which will be routed to the client, resulting in the connection broken down at the client side. 

\Major{For the SSH DoS attack, we define the result that the client or the server cannot receive messages from each other as a successful attack,} 
which includes two cases. First, after the attacker receives the \texttt{ACK} packet from the server, it can send a TCP \texttt{RST} packet with the exact sequence number to the server, resulting in the connection \Major{terminated} at the server side. Second, as the attacker has replaced the NAT mapping at the router, the source port of the packet will be translated if it happens that the client sends a packet to the server at this stage. When the packet arrives at the server, it will incur a \texttt{RST} packet as there is no corresponding connection, which will be routed to the client, resulting in the connection \Major{terminated} at the client side. 
\Major{In the context of the FTP hijacking attack, the attack can be deemed successful when the attacker manages to download files from the FTP server that belong to the victim client.}
\Major{And for the HTTP injection attack,} we define the result that the client receives forged packets, and the falsified data is displayed on the web page as a successful attack. Compared with \Major{SSH DoS and FTP hijacking} attacks, the conditions are more difficult to meet. As the attacker only knows that the request interval is 60 seconds for the \textit{www.{ANONYMOUS}.com} website while it does not know when the client will request an update. The client may request new data during the attack, which results in the connection being terminated, and we strictly take this case as a failure.

We repeat the experiments 20 times in each tested Wi-Fi network. \Major{Each experiment is conducted independently with a renewed connection between the client and server.} In order to limit the time of experiments, we take an experiment as a failure if the attacker cannot terminate the connection, \Major{download private files, or} the forged data does not show up on the client's web page in 5 minutes.
As mentioned before, it takes time (mostly 1 second or 10 seconds) for the mapping to disappear completely. The countdown will be refreshed if the client sends packets during this period, which may interfere with the time cost and success rate of our attack.
% , although the attacker can repeat the attack continuously. 
%
To simulate real-world situations, we require the tested client to send requests to the server for random times, and we set the interval between two requests as a random number from 5 to 30 seconds during the 5 minutes of an experiment. We will further investigate the impacts of communication intervals between the client and server and the timeout of NAT mappings in Section \ref{sec:practical_consideration}.

\noindent \textbf{Experimental Results}. 
% We evaluate our attack against 93 real-world Wi-Fi networks to cover the most typical public Wi-Fi scenarios, e.g., Wi-Fi networks in coffee shops, hotels, shopping malls, airports, campuses, and office buildings. The experimental results illustrate that more than 81\% of the real-world Wi-Fi networks (i.e., 75 out of the 93 evaluated networks) are vulnerable that they satisfy all of the conditions of our attacks. 
We evaluate our attack against 93 real-world Wi-Fi networks to cover the most typical Wi-Fi scenarios, e.g., Wi-Fi networks in coffee shops, hotels, shopping malls, campuses, and office buildings. 
As the attacker can sniff the non-encrypted packets on the air in open networks directly, we mainly launch our experiments under networks protected by WPAs from home mode networks and enterprise mode networks, e.g., 45 with WPA2-Personal enabled (home mode), 22 with WPA2-Enterprise enabled (enterprise mode), and 26 with WPA3-Personal enabled (home mode) and we do not find any network with WPA3-Enterprise enabled. 
The experimental results illustrate that more than 81\% of the real-world Wi-Fi networks (i.e., 75 out of the 93 evaluated networks) are fully vulnerable that they satisfy all of the conditions of our attacks. \Major{For the other 18 Wi-Fi networks, 9 of them have AP isolation enabled, which prevents the detection of potential victims and thwarts the HTTP injection attack. However, the SSH DoS and FTP hijacking attacks remain unaffected. Our attack fails in 7 networks as they do not use the vulnerable routers, and we cannot get the router's external IP as described in Section \ref{sec:network_probe} in the rest two networks.}
% , which allow an attacker to hijack victim clients' TCP connections to a given server, thus causing a denial of service attack or a privacy breach in the real world. 
\Major{We successfully acquire the external IP addresses of routers using the route-recording method in 80 networks, involving router models from 22 vendors (i.e., Ubiquiti, Amazon, Google, Tenda, ASUS, Netgear, Huawei, Linksys, Xiaomi, Ruijie, ZTE, H3C, Wavlink, Comfast, IP-COM, Skyworth, Netcore, iKuai, WiMaster, GL.iNet, JdCloud, and China Mobile). Additionally, we employ the scanning method in 11 networks, utilizing router models from 6 vendors (i.e., D-Link, Volans, pfSense, and some models of 360, Mercury, and TP-Link). In the remaining two networks using the router models from Cisco Meraki and Aruba, we fail to get the external IP address unless we log in to the control page of the router with the help of the network administrators, and we take these networks as failures.}

\begin{table}[h]
\small
\renewcommand\arraystretch{1.2}
\setlength\tabcolsep{1pt}
\centering
\caption{Experimental results in our tests (on average).}
\begin{threeparttable}
\scalebox{0.85}{
\begin{tabular}{ccccccc} 
\toprule

\multicolumn{1}{c}{\begin{tabular}[c]{@{}c@{}}\textbf{Attack}\\\textbf{Type}\end{tabular}} & \multicolumn{1}{c}{\begin{tabular}[c]{@{}c@{}}\textbf{Inferring}\\\textbf{Port(s)}\end{tabular}} & \multicolumn{1}{c}{\begin{tabular}[c]{@{}c@{}}\textbf{Getting}\\\textbf{SEQ/ACK(s)}\end{tabular}} & \multicolumn{1}{c}{\begin{tabular}[c]{@{}c@{}}\textbf{Finishing}\\\textbf{Attacking(s)}\end{tabular}} & \multicolumn{1}{c}{\begin{tabular}[c]{@{}c@{}}\textbf{Total}\\\textbf{Time(s)}\end{tabular}} & \multicolumn{1}{c}{\begin{tabular}[c]{@{}c@{}}\textbf{BW}\\\textbf{(pkts) }\end{tabular}} & 
\multicolumn{1}{c}{\begin{tabular}[c]{@{}c@{}}\textbf{Success}\\\textbf{Rate}\end{tabular}} \\ 
\hline

% \rowcolor[rgb]{0.871,0.871,0.871} 
SSH DoS          &8.1  & 8.4  & 1.0 & 17.5 &4000 & 87.4\% \\
FTP Hijacking    &9.1  & 9.2  & 1.1 & 19.4 &4000 & 82.6\% \\
HTTP Injection   &9.4  & 15.2 & 29.9 & 54.5 &4000 & 76.1\% \\
\bottomrule

\end{tabular}}
% \begin{tablenotes}
%        \footnotesize
%        \item[] * means the total time of a successful attack, i.e., breaking down SSH or injecting forged data to the web page.
%     \end{tablenotes}
\end{threeparttable}
\label{attack_results}
\arrayrulecolor{black}
\end{table}
\vspace{-1mm}

Next, we elaborate on our experimental results in the Wi-Fi networks as shown in Table~\ref{attack_results}. In the case of the SSH DoS attack, the average time cost of identifying the client's source port is 8.1 seconds with a bandwidth of 4000 packets per second, which is much shorter than previous methods \cite{ccsfeng,chen2018off,cao2016off} as we only need to transfer packets in the same LAN and we are not restricted by rate limits. And the average time cost of obtaining the exact sequence and acknowledgment numbers is 8.4 seconds, as this step mainly relies on the default settings of the timeout of \texttt{CLOSE} state in NAT mappings. Besides, the communication between the client and the server may also influence the time cost. Finally, the average time cost of totally \Major{terminating} an SSH connection is 17.5 seconds, and the average success rate is 87.4\%. 
\Major{The failure cases in the tests are due to continuous communications between the client and the server (e.g., the client requests a file download and related packets will always refresh the NAT mapping).}
% The failure cases in the 20 tests are mainly due to the time limit of our requirement. We find that in these cases the client continuously communicates with the server or requires a file download within the tested 5 minutes. 
After the attack succeeds, the client's SSH terminal will be stuck for a period of time, which greatly affects the user experience. 
% Note that the DoS attack is also applicable to compromising applications secured by other encrypted traffic (e.g., HTTPS). 

\Major{For the FTP hijacking attack, the average time costs of identifying the client's source port and getting the sequence and acknowledgment numbers are 9.1 and 9.2 seconds, respectively, which results in a time cost of 19.4 seconds for the entire attack to get a private file from the server with a success rate of 82.6\% on average. The failure cases in the tests are due to two reasons. The first is the same as the cases in the SSH DoS attack, i.e., continuous communications. The second is that the attacker happens to begin the attack when the victim connection has been constructed while the victim has not logged in. 
}

\begin{figure}[h]
	% \vspace{-2mm}
	\begin{center}
            \vspace{7pt}	%修改为7pt
		\includegraphics[width=\linewidth]{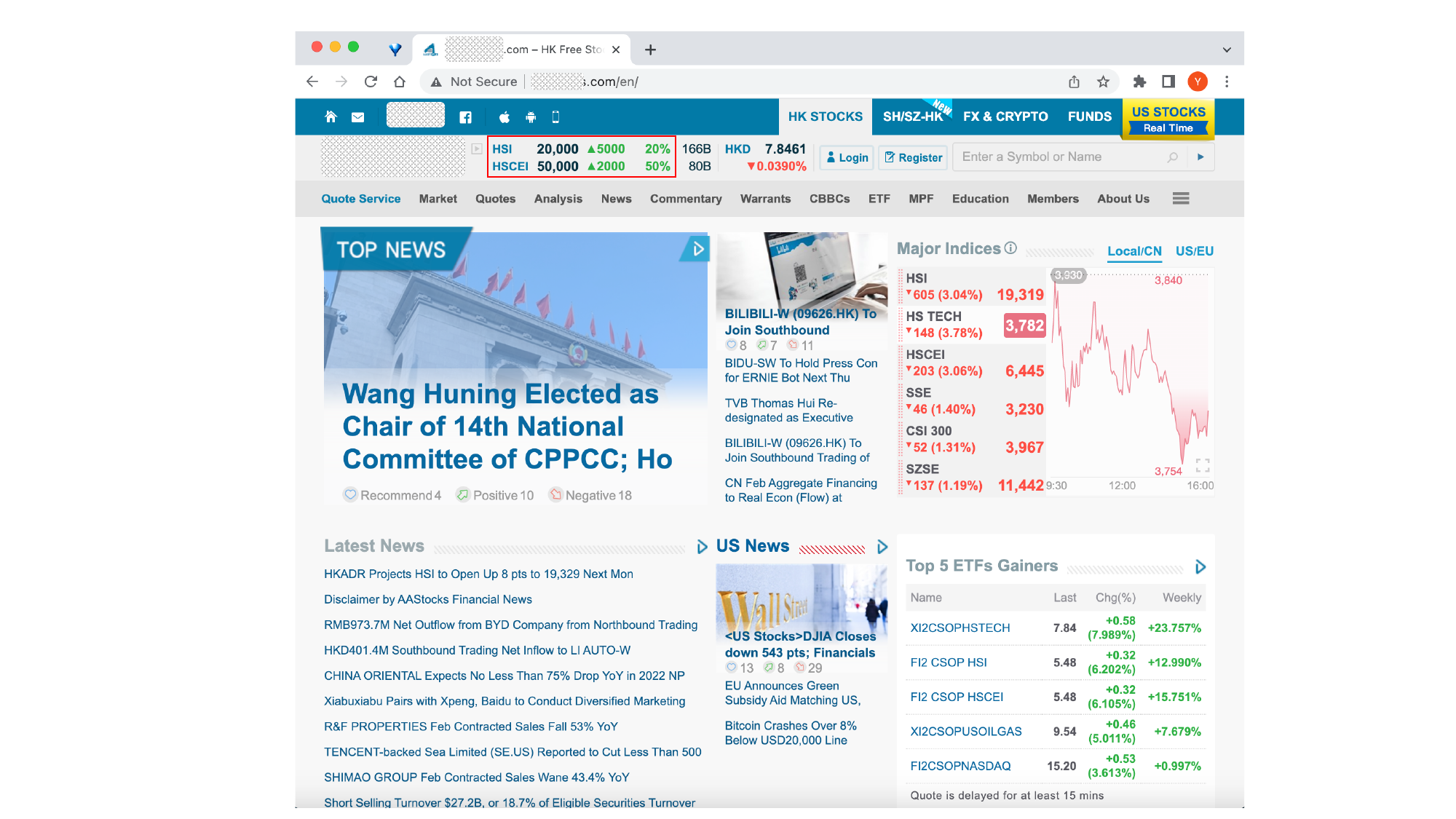}
		% \vspace{2mm}
		\caption{Snapshots of web poisoning}
		\label{aastocks}
	\end{center}
	% \vspace{-4mm}
\end{figure}

As for the \Major{HTTP injection attack}, the average time costs of identifying the client's source port, obtaining the sequence and acknowledgment numbers, and the whole attack to inject forged data to the victim's web page are 9.4 seconds, 15.2 seconds, and 54.5 seconds, respectively,
% Note that we take the time for the reconstruction of the original NAT mapping into account.
with an average success rate of 76.1\%.
Compared with the SSH DoS and FTP hijacking attacks, it is more time-consuming as the attacker has to wait for more time to inject fake responses until the next request, and it has a lower success rate 
% as the attacker has to complete the above attack and inject a fake message at the right time, or the connection may be terminated. 
\Major{as the connection may be terminated if the client sends a request when the NAT mapping has been occupied by the attacker. Another scenario of failure occurs when the attacker fails to win the condition race of returning responses with the server, i.e., the client accepts the right data from the server.}
Figure \ref{aastocks} shows the snapshot of our HTTP injection attack against \textit{www.{ANONYMOUS}.com}. The original website shows that the \texttt{HSI} number is 19,319, and it has reduced by 605 with a drop rate of 3.04\%. After the attack, the victim will find that the \texttt{HSI} number is 20,000 and it has increased by 5000 with a growth rate of 20\%. The same is true for the data of \texttt{HSCEI}. The attack may lead to wrong stock purchase or sale, affecting the financial status of the victim. 

\begin{table*}[!t]
% \vspace{-110mm}
\small
\renewcommand\arraystretch{1.3}
\setlength\tabcolsep{10pt}
\centering
\caption{Experimental results of TCP Attacks in 30 Wi-Fi networks.}
\label{measurement-results}
%\resizebox{\linewidth}{!}{%
\scalebox{0.8}{
\begin{tabular}{p{1cm}<{\centering}ccccccccc} 
\toprule

\textbf{No.} 
& \textbf{\tabincell{c}{Network \\Mode}} 
& \textbf{SSID} 
& \textbf{\tabincell{c}{Router \\Vendor}}   
& \textbf{\tabincell{c}{Wi-Fi \\Generation}}
& \textbf{\tabincell{c}{WPA2/3 \\Enterprise/Personal}}
& \textbf{\tabincell{c}{Attack \\Result}}
& \textbf{\tabincell{c}{Time \\Cost (s)}}
& \textbf{\tabincell{c}{Success \\Rate}} \\ 

\hline

1 & Enterprise mode & Campus 1 & Huawei & Wi-Fi 6 & WPA2-Enterprise & SSH DoS & 15.43 & 18/20 \\

2 & Enterprise mode & Campus 2 & TP-Link & Wi-Fi 4 & WPA2-Enterprise & FTP Hijacking & 10.32 & 18/20 \\

3 & Enterprise mode & Campus 3 & H3C  & Wi-Fi 6 & WPA2-Enterprise & HTTP Injection & 48.87 &  15/20 \\
\hline

4 & Enterprise mode & Enterprise 1 & TP-Link  & Wi-Fi 6 & WPA2-Enterprise & SSH DoS & 11.56 & 16/20 \\

5 & Enterprise mode & Enterprise 2 & TP-Link  & Wi-Fi 5 & WPA2-Enterprise & FTP Hijacking & 11.43 & 18/20 \\

6 & Enterprise mode & Enterprise 3 & Netcore  & Wi-Fi 6 & WPA2-Enterprise & HTTP Injection & 87.20 & 15/20 \\

\hline

7 & Enterprise mode & Office building 1 & TP-Link & Wi-Fi 5 & WPA2-Enterprise & SSH DoS & 9.56 & 18/20 \\

8 & Enterprise mode & Office building 2 & iKuai  & Wi-Fi 6 & WPA2-Enterprise & FTP Hijacking & 21.46 & 17/20 \\

9 & Enterprise mode & Office building 3  & Mercury  & Wi-Fi 6 & WPA2-Enterprise & HTTP Injection & 31.14 & 15/20 \\
\hline

10 & Enterprise mode & Hotel 1 & Netcore  & Wi-Fi 5 & WPA2-Enterprise & SSH DoS  & 15.75 & 18/20 \\

11 & Enterprise mode & Hotel 2 & D-Link  & Wi-Fi 6 & WPA2-Enterprise & FTP Hijacking & 9.45 & 19/20 \\

12 & Enterprise mode & Hotel 2 & iKuai  & Wi-Fi 6 & WPA2-Enterprise & HTTP Injection & 71.32 & 16/20 \\
\hline

\rowcolor[rgb]{0.871,0.871,0.871} 13 & Home mode &  Restaurant 1  & TP-Link & Wi-Fi 5 & WPA2-Personal & SSH DoS & 8.95 & 17/20 \\

\rowcolor[rgb]{0.871,0.871,0.871} 14 & Home mode &  Restaurant 2  & Comfast  & Wi-Fi 5 & WPA2-Personal & FTP Hijacking & 21.56 & 18/20 \\

\rowcolor[rgb]{0.871,0.871,0.871} 15 & Home mode & Restaurant 3 & Skyworth  & Wi-Fi 6 & WPA2-Personal & HTTP Injection & 62.35 & 13/20 \\
\hline

\rowcolor[rgb]{0.871,0.871,0.871} 16 & Home mode & Coffee shop 1 & Mercury & Wi-Fi 4 & WPA2-Personal & SSH DoS & 8.98 &  17/20 \\

\rowcolor[rgb]{0.871,0.871,0.871} 17 & Home mode & Coffee shop 2 & TP-Link & Wi-Fi 4  & WPA2-Personal & FTP Hijacking & 9.29 & 18/20\\

\rowcolor[rgb]{0.871,0.871,0.871} 18 & Home mode & Coffee shop 3 & Wavlink & Wi-Fi 5  & WPA2-Personal & HTTP Injection& 45.22 & 13/20 \\

\hline

\rowcolor[rgb]{0.871,0.871,0.871} 19 & Home mode & Shopping mall 1 & Tenda  & Wi-Fi 6 & WPA3-Personal & SSH DoS & 24.23 & 18/20 \\

\rowcolor[rgb]{0.871,0.871,0.871} 20 & Home mode & Shopping mall 2 & TP-Link  & Wi-Fi 4 & WPA2-Personal & FTP Hijacking & 11.44 & 19/20 \\

\rowcolor[rgb]{0.871,0.871,0.871} 21 & Home mode & Shopping mall 3 & Huawei  & Wi-Fi 6 & WPA3-Personal & HTTP Injection & 78.44 & 15/20\\
\hline

\rowcolor[rgb]{0.871,0.871,0.871} 22 & Home mode & Bookstore 1 & 360 & Wi-Fi 5 & WPA2-Personal & SSH DoS & 19.45 & 18/20 \\

\rowcolor[rgb]{0.871,0.871,0.871} 23 & Home mode & Bookstore 2  & Xiaomi & Wi-Fi 6 & WPA3-Personal & FTP Hijacking & 10.61 & 18/20 \\

\rowcolor[rgb]{0.871,0.871,0.871} 24 & Home mode & Bookstore 3 & H3C & Wi-Fi 6 & WPA3-Personal & HTTP Injection & 56.12 & 14/20\\
\hline

\rowcolor[rgb]{0.871,0.871,0.871} 25 & Home mode & Experience store 1 & Xiaomi & Wi-Fi 6 & WPA3-Personal & SSH DoS & 16.97 & 17/20\\

\rowcolor[rgb]{0.871,0.871,0.871} 26 & Home mode & Experience store 2 & Huawei  & Wi-Fi 6 & WPA3-Personal & FTP Hijacking & 23.98 & 18/20 \\

\rowcolor[rgb]{0.871,0.871,0.871} 27 & Home mode & Experience store 3 & Xiaomi & Wi-Fi 5 & WPA2-Personal & HTTP Injection & 52.14 & 16/20 \\

\hline

\rowcolor[rgb]{0.871,0.871,0.871} 28 & Home mode &  Cinema 1 & Ruijie & Wi-Fi 5 & WPA2-Personal & SSH DoS & 8.89 & 19/20 \\

\rowcolor[rgb]{0.871,0.871,0.871} 29 & Home mode &  Cinema 2 & Mercury  & Wi-Fi 6 & WPA3-Personal & FTP Hijacking & 11.31 & 18/20\\

\rowcolor[rgb]{0.871,0.871,0.871} 30 & Home mode &  Cinema 2 & Huawei  & Wi-Fi 6 & WPA3-Personal & HTTP Injection & 54.26 & 16/20\\

% 29 &  Library 1 & Tenda & Wi-Fi 6 & SSH DoS & 22.64 & 19/20 \\
% 30 &  Library 2 & Linksys  & Wi-Fi 6 & Web Manipulation & 47.76 & 16/20\\

\bottomrule
\end{tabular}
}
%\vspace{-4mm}
\end{table*}

The experimental results of 30 Wi-Fi networks in our investigations are listed in Table~\ref{measurement-results}. As shown in the first row, an enterprise mode Wi-Fi network with the SSID of ``Campus 1\footnote{We anonymized the real SSIDs of the Wi-Fi networks due to ethical considerations.}'' is located in a campus. The router is produced by the vendor of Huawei, and it supports the generation of Wi-Fi 6. We take the experiment of SSH DoS attack in this network, and it takes the attacker 15.43 seconds to \Major{terminate} an SSH connection with a success rate of 90\%. 
% Factors leading to the failures of our experiments in the evaluated networks are detailed in \ref{sec:practical_consideration}.

%-------------------------------------------------------------------------------
\section{Discussion}\label{sec:discussion}
In this section, we discuss the factors that affect the attack's effectiveness. \Major{We also compare our attack with existing attacks in WLANs. Besides,} we extend our attack model to launch a remote TCP DoS attack from an attacker on the Internet.

\subsection{Factors Impacting the Attack}\label{sec:practical_consideration}

\noindent \Major{\textbf{Impacts of Traffic Load.}}
% As mentioned in \ref{sec:hijack}, the NAT mappings will be cleaned completely after their timeouts when receiving TCP \texttt{RST} packets. In our investigation, the default settings of the vulnerable routers are 1 second or 10 seconds, during which the timeout will be refreshed if related packets are traveling through the router. In this way, the attack will be interfered with if the client communicates with the server frequently. We add experiments to investigate their influences. 
\Major{
We analyze the impact of traffic load on the attacks from two aspects: 1) the bandwidth between the client and the server, and 2) the communication interval between the client and the server.}
\Major{
First, we extend the experiments of the FTP hijacking attack with varied bandwidths (i.e., 10KBps, 100KBps, 1000KBps) and set the communication interval to 16 seconds with a NAT mapping timeout of 10 seconds. We repeat the experiment 50 times for each bandwidth and record the time cost and the number of successful attacks. The experimental results show that the average time costs (i.e., 29.46 seconds, 28.78 seconds, 29.21 seconds) and success rates (i.e., 96\%, 94\%, 94\%) remain largely unaffected since our attack mainly relies on the time interval left for the attacker to clean the NAT mappings.}

\begin{figure*}[t]
    \centering
    \begin{minipage}[t]{\linewidth}
        \subfigure[The time costs and success rates in different communication intervals when the timeout of the NAT mapping is 1 second.]{
            \begin{minipage}[t]{\linewidth}
                \centering
                \includegraphics[width =\textwidth]{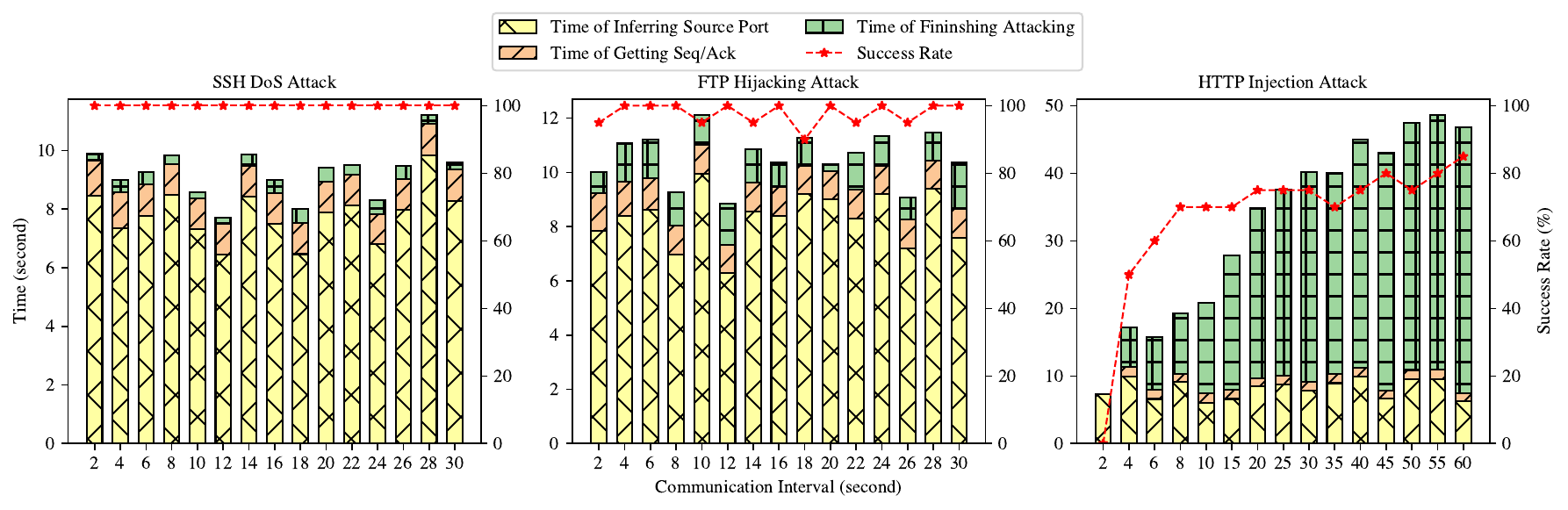}
                \label{fig:time-1}
            \end{minipage}}
        \subfigure[The time costs and success rates in different communication intervals when the timeout of the NAT mapping is 10 seconds.]{
            \begin{minipage}[t]{\linewidth}
                \centering
                \includegraphics[width =\textwidth]{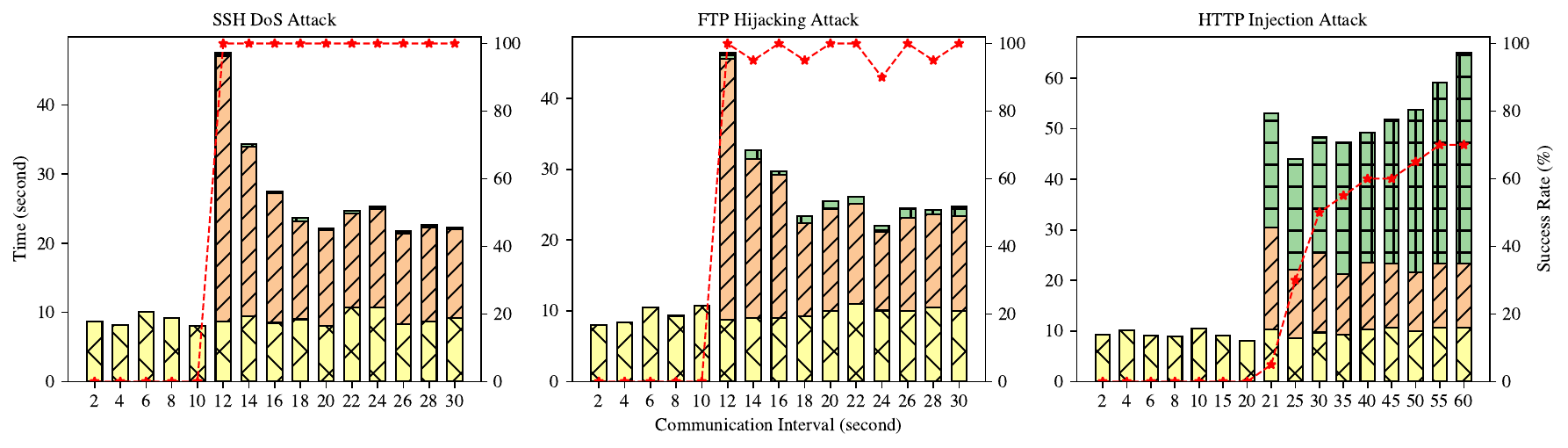}
                \label{fig:time-10}
            \end{minipage}}
    \end{minipage}
    \caption{The time costs and success rates in different communication intervals and different NAT mapping timeouts.}
    \label{fig:time}
\end{figure*}

\Major{
Second, we evaluate the three attacks under various communication intervals (e.g., 2 seconds, 4 seconds, 6 seconds, etc.) between the client and server, with a bandwidth of 100KBps. We repeat the experiments 20 times for each communication interval and record the time used in each attack phase and count the successful attacks.
The experimental results are shown in Figure \ref{fig:time}. 
}

\Major{
Here, we take the FTP hijacking attack as an example. As shown in Figure \ref{fig:time-1}, when the NAT mapping timeout is set to 1 second, if the communication interval is below 1 second, the attack will fail due to the continuously refreshed NAT mappings. When the communication interval is above 1 second, the attack can succeed with a high success rate (97.67\%), where the small partial failures are due to the attacks being launched during the login phase of the FTP application. The average time cost is less affected by the communication interval, and it shows a fluctuating trend, which mainly depends on the time to infer the client's source port. 
Similarly, when the NAT mapping timeout is 10 seconds (as shown in Figure \ref{fig:time-10}), if the communication interval is below 10 seconds, the attack will fail due to the same reason. When the interval is above 10 seconds, the attack can succeed with a high success rate (97.5\%). The average time cost shows a downward trend with the increment of communication interval, as the attacker can try fewer times and wait less time for the NAT mappings to be cleaned when the communication interval is longer in an attack.
}

\noindent \NewMajor{\textbf{Distribution of Time Cost and Failures.}}
\NewMajor{
%According to the experiments above, apart from 
Besides evaluating the FTP hijacking attack, we also measure the time costs and failure reasons of the other two attacks, i.e., the SSH DoS attack and the HTTP injection attack.
For the SSH DoS attack, as shown in Figure \ref{fig:time-1}, when the NAT mapping timeout is 1 second, if the communication interval is below 1 second, the attack will fail. However, when the communication interval is larger than 1 second, the attack can always succeed. The average time cost is not very impacted by the communication interval, and it shows a slight fluctuating trend with an average value of 9.24 seconds. The main cost is incurred by inferring the client's source ports. Figure \ref{fig:time-10} shows similar results when the NAT mapping timeout is 10 seconds. 
}

\NewMajor{
As for the HTTP injection attack, as shown in Figure \ref{fig:time-1}, 
when the NAT mapping timeout is 1 second, if the communication interval is below 2 seconds, the attack will fail due to the continuously refreshed NAT mappings, or the connection will be terminated during the reconstruction of the original NAT mapping as stated in Section \ref{sec:Attack_Evaluation}. 
However, when the communication interval is larger than 2 seconds, with the increase of the communication interval, the time cost and success rate tend to increase. The time cost mainly depends on the waiting period until the client sends a request, as injections before that will not be accepted. In this way, the longer the communication interval is, the more waiting time will cost on average. The failures mainly come from two aspects. First, the original connection may also be terminated. Second, the attacker should inject data before the response of the server when the victim client sends a new request. 
Similar results can be found in Figure \ref{fig:time-10} when the NAT mapping timeout is 10 seconds. The difference is that the time to get sequence and acknowledgment numbers occupies a larger proportion compared with that of 1 second.
}

\noindent \NewMajor{\textbf{Impacts of NAT Mapping Timeout.}
We analyze the impact of NAT mapping timeout by comparing the experimental result of the HTTP injection attack when we set the communication interval between the victim client and the server to 60 seconds, reflecting an actual client-server communication scenario in the real world. %Subsequently, we perform a in-depth analysis of the impact of different NAT mapping timeouts on our attack.
%
% We consider the two NAT mapping timeouts (i.e., 1 second and 10 seconds), which are the only two timeouts commonly used in 66 of the 67 tested router models.
%
When setting the NAT mapping timeout to 1 second, we have a time cost of 46.80 seconds and a success rate of 85\%. However, we observe a time cost of 64.96 seconds and a success rate of 70\% when the NAT mapping timeout is 10 seconds. 
}
\NewMajor{
The reason is that a larger timeout value incurs longer probing overhead. For example, compared to the cases with a 1-second timeout, when the timeout is 10 seconds, the time to obtain sequence and acknowledgment numbers takes up a larger proportion as the attacker must wait for the mappings to be completely cleared (i.e., at least 10 seconds). Under a long timeout, the connection is more likely to be unintentionally terminated, and thus the success rate is reduced. In summary, the increase of NAT mapping timeouts will incur increased time costs and decreased success rates.  
}

\noindent \textbf{Impacts of AP Isolation.}
% Specific network configurations enforced by the target Wi-Fi networks (e.g., AP isolation) may influence the results of our attack. For example, in our measurement study, we discover 9 Wi-Fi networks where internal clients in the same network cannot communicate with each other. \Major{TODO::TCP DoS and hijacking attacks are not influenced. Injection attack may be influenced. add some experimental results.} With the policy enabled, it's hard for the attacker to probe potential victim clients in the first phase of the attack and it cannot reconstruct the original NAT mapping at the router by sending spoofed TCP \texttt{ACK} packets to all of the potential victim clients. This AP isolation policy throttles our attack while it may affect network interoperability. Besides, only less than 10\% (9 out of 93) of the real-world public Wi-Fi networks enforce AP isolation in our test.
\Major{AP isolation may influence some phases of our attack. With the policy enabled, the attacker cannot probe potential victim clients in the first phase of the attack, and it cannot reconstruct the original NAT mapping at the router by sending spoofed TCP \texttt{ACK} packets to all of the potential victim clients when launching the HTTP injection attack. However, the SSH DoS and FTP hijacking attacks are not affected as the attacker does not need to send packets directly to the victim client. 
Besides, only less than 10\% (9 out of 93) of real-world Wi-Fi networks we observed enforce AP isolation.
We also enabled AP isolation on three routers (i.e., TP-Link TL-XDR6020, Linksys E5600, and Xiaomi RA81) in our laboratory and performed the two attacks. 
% Fundamentally, the vulnerability is rooted in the router's NAT behaviors.
% This AP isolation policy may throttle our attack while it may affect network interoperability. 
The experiment results show that the time cost and success rate are not affected by AP isolation. 
}
%Major revision
\subsection{Comparison with Prior Attacks in Wi-Fi Networks}\label{sec:attack_comparison}
% Ever since Wi-Fi technology was invented, numerous attacks have been discovered, such as ARP poisoning, eavesdropping and Evil-Twin attacks.
%
\noindent 
\Major{\textbf{ARP Poisoning Attack.}
Compared with our attack, a successful ARP poisoning attack can intercept traffic in both directions, i.e., from the victim client and the router, while our attack can only intercept TCP traffic from the router. However, ARP poisoning attack in wired or wireless LANs has been well-researched since it appeared. Users can install some open-sourced tools \cite{ARP-AntiSpoofer, shARP, 360-ARP} to prevent the attack. Besides, some routers (e.g., TP-Link) offer built-in ARP protections. 
% Users can install some open-sourced tools \cite{ARP-AntiSpoofer, shARP, linux-ARP, 360-ARP} to prevent the attack and some routers (e.g., TP-Link) offer built-in ARP protection. We test the ARP poisoning attack and its prevention on a PC with ~\cite{360-ARP} under the TP-Link router and we find that only both with the protection of routers as well as tools on client devices, the ARP poisoning attack can be fully prevented. 
Moreover, AP isolation can also defend against the ARP poisoning attack effectively by preventing communication between clients, while our attack is only partially affected as stated in Section \ref{sec:hijack} and Section \ref{sec:practical_consideration}.
We make a further empirical study in 10 real-world Wi-Fi networks. Three of them have enabled AP isolation and can prevent the ARP poisoning attack. On the other Wi-Fi networks, ARP poisoning can succeed. However, we observe that it fails when we enable protections on routers and client devices (e.g., using the tool developed in \cite{360-ARP}).
In contrast, our attack can still succeed in these networks, even when these protections are in place.
}

\noindent
\Major{\textbf{Eavesdropping Attack.}
Against WPA2-Personal mode Wi-Fi networks, a malicious attacker who knows the pre-shared key can sniff the frames in the air of other clients. If it wants to decrypt the frames, it needs to capture the 4-way handshake frames when other clients are connecting to the network. Though the attacker can force the victim client to be detached from the current AP by sending fake deauthentication frames~\cite{vanhoef22deauth} and wait for its re-connection, the attack is perceivable by the victim that its device will lose Internet access. However, our attack is stealthier as the attacker only needs to connect to the same network and does not need to make the client disconnect from the existing network to launch the attack. Besides, it's much harder to decrypt the frames encrypted with WPA2-Enterprise mode or WPA3-Personal mode, while our attack can also influence these networks.
}

\noindent 
\NewMajor{\textbf{Rogue AP Attack.}
The malicious insider who knows the pre-shared key can also create a rogue clone (evil twin) of the network and entice unsuspecting victims to connect to it, thus intercepting all the traffic~\cite{rogueAP-attack,evil-AP}. This attack requires broadcasting the same SSID, which can be detected by the network administrator, and some routers also provide protection strategies such as Rogue AP detection \cite{Huawei-rogueAP,Linksys-rogueAP}. In contrast, our attack is stealthier as there is no specific strategy provided to detect our attack. In addition, a lightweight device compromised by the attacker remotely may not have enough resources to provide the services as a rogue AP. The attacker who is physically in the LAN can set its own device as a rogue AP, but it has to provide a stronger signal than the original AP, and thus the influence is limited to clients in close proximity. Conversely, our attack does not face the signal race that any device in the same Wi-Fi network can launch the attack and potentially influence all clients.
Besides, enterprise mode Wi-Fi networks can protect clients from the Rogue AP attack. However, our empirical measurements have revealed that our attack can compromise the traffic of clients within 22 different enterprise mode Wi-Fi networks.
}

\NewMajor{
Compared with the prior works, our attack leverages a new side channel vulnerability of the NAT behaviors in routers that can be exploited to hijack TCP connections by off-path attackers, even in enterprise mode networks with AP isolation.
Moreover, lots of strategies have been proposed to prevent prior attacks, while our attack is a novel one whose vulnerability has existed in routers for years. Additionally, our attack serves as a valuable supplementary attack in networks equipped with defense measures against existing attacks.
% And till now, some vendors have provided enough feedback that they are repairing their products with some countermeasures while others are still silent that we hope with this paper they could pay more attention to this vulnerability to fix it and protect the Wi-Fi users. 
}

% It is not a MITM attack
% \subsection{Attacks of Man-in-the-Middle}\label{sec:mitm_discussion}
% As stated in ~\ref{sec:hijack}, after the attacker has obtained the exact sequence and acknowledgment numbers from the server by evicting the NAT mapping of the client and constructing a new one of itself at the router, it can impersonate the client to request data from the server. With the translation of IP addresses at the router, the request will match the associated connection at the server and it will return the exact responses back to the attacker instead of the victim client anymore. We take the following case as an example of this attack. The victim client first logins to the server and requests private information from it. After that, the attacker can bypass the initial verification stage by replacing the victim client as a MitM to send requests to the server and steal the victim's confidential information, which may lead to permission bypass and privacy leakage. We will deeply dive into this kind of attack and how to make it well applicable to real-world applications is what we may consider in the future.

% Besides, if the victim client intends to communicate with the server during this phase, the source port of its request will be translated and it will incur the server to respond with a \texttt{RST} packet back. Thus the connection will be broken down at the client side.

\subsection{Extending the Attack Model}\label{sec:modes}
In our extended model, we eliminate the requirement that the attacker and the victim client have to be located in the same Wi-Fi network. Instead, we demonstrate that a remote attacker from the Internet can launch a DoS attack on TCP connections between victim clients behind a vulnerable router and an external server. We require that the attacker can send packets with spoofed source IP addresses, which is a practical assumption considering that approximately a quarter of autonomous systems still do not employ source address validation (SAV), as reported by the Spoofer project \cite{SAV-deployment}.

\begin{figure}[h]
	% \vspace{-2mm}
	\begin{center}
     % \vspace{7pt}	%修改为7pt
	  % \setlength{\abovecaptionskip}{0mm} %调整图片标题与图距离
   %    \setlength{\belowcaptionskip}{-7mm}
		\includegraphics[width=\linewidth]{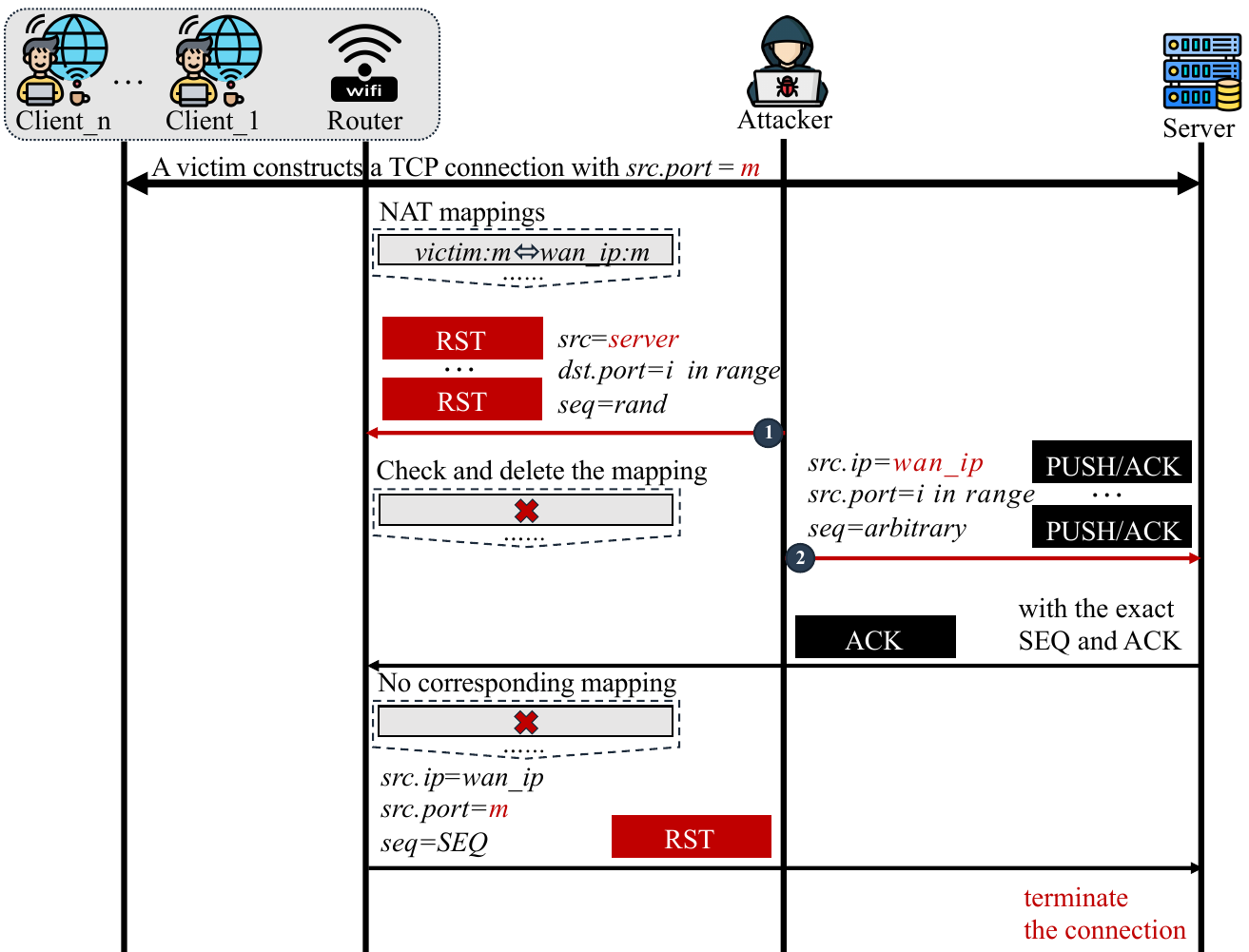}
		% \vspace{2mm}
		\caption{Remote TCP DoS attack}
		\label{TCPDoS}
	\end{center}
	% \vspace{-3mm}
\end{figure}

Assuming that there is a live connection between an outside server and the victim client who resides behind a vulnerable router. Compared to the original attack model, the attacker cannot infer the source port of the victim client anymore using the method before. However, 
as we show in Figure \ref{TCPDoS}, 
the attacker can send forged TCP \texttt{RST} packets covering the entire space of possible source ports to the public IP address of the vulnerable router. As the routers do not check the sequence number specified in TCP packets strictly, these \texttt{RST} packets can easily bypass routers' checks to clean the possible NAT mapping of the victim connection.
% After the NAT mapping disappears, the attacker can spoof as the router to send forged TCP data packets to the server, covering the entire space of possible source ports, too. The server will respond to the matched one with an \texttt{ACK} packet specified with the current sequence and acknowledgment numbers to the router. 
After the NAT mapping disappears, \Major{the attacker can send forged TCP data packets to the server with a spoofed source IP address of the router's public IP}, covering the entire space of possible source ports, too. The server will respond to the matched one with an \texttt{ACK} packet specified with the current sequence and acknowledgment numbers to the router. 
However, as there is no corresponding NAT mapping of this \texttt{ACK} packet anymore, the router will just send a \texttt{RST} packet back to the server with the sequence number received just now. Then the connection will be terminated from the server side. If the client continues communicating with the server, it will receive a TCP \texttt{RST} packet back from the server afterward. In this way, the attacker can interfere with TCP communications between the victim and server, causing a DoS attack and affecting user experience.

The attacker needs to detect such victim clients who access the Internet through these vulnerable routers. 
% Some prior works \cite{findNAT1,findNAT2} have proposed methods on how to identify NAT networks and probe clients behind them. 
We find that it is of great convenience for the attacker to identify these vulnerable routers through open search engines \cite{shodan,fofa}, which contain a large amount of publicly accessible devices (e.g., routers, web servers, and webcams). For example, millions of TP-Link routers with public IP addresses can be found through FOFA \cite{fofa}. We estimate that there are tens of millions of vulnerable routers existing in the world which may be influenced by the attack, and we believe this attack is promising and practical and may affect many more users. We leave it as future work to validate the real-world impact in practice owing to reasons such as ethical considerations.

%-------------------------------------------------------------------------------
\section{Countermeasures}\label{sec:countermeasures}
\noindent \textbf{Responsible Vulnerability Disclosure.}
We have reported the issue to the affected manufacturers by submitting vulnerability reports and contacting them via email. At the time of writing, we have received positive responses from the OpenWrt community that confirms our findings and has released patches to fix the vulnerability, and seven router vendors (i.e., TP-Link, Huawei, Xiaomi, 360, Mercury, \Major{Ubiquiti}, and Linksys) that have all acknowledged our reports and are trying to repair their products.
\Major{In addition, we have been assigned 10 CVE numbers for the vulnerability in different vendors (i.e., TP-Link, Linksys, Mercury, Ruijie, D-Link, Comfast, H3C, OpenWrt, Wavlink, and 360)}. 
The other vendors are still investigating the vulnerability. 
We also provide them with countermeasure suggestions to mitigate the identified attack, and some of them have been adopted by the vendors. As mentioned in Section \ref{sec:threat_model}, we outline several conditions that characterize a vulnerable router implementation. Intuitively, any breach of these conditions will render the attack ineffective. 

\noindent \textbf{Random Port Allocation}. The first solution is for the router to use the \textit{random selection} strategy when creating new NAT mappings. In detail, the router can choose a random port from the available port pool and record the port translation when allocating new mappings. 
% As there is no specific requirement of whether to preserve the original source port or not in RFCs \cite{rfc5382}, the TCP packets can also be forwarded normally with this strategy. 
% For example, in our attribute test of routers (see Table \ref{all-routers}), we find that one enterprise wired router model ``AR6140E-9G-2AC'' from Huawei adopts the \textit{random selection} strategy by default.
With this strategy, the attacker cannot identify whether the port has been used by other internal hosts, and the attack will be foiled. 
\Major{It should be noted that some TCP punch-through schemes (e.g., TCP simultaneous open) may be influenced by \textit{random selection} as they rely on port prediction~\cite{NATHole-SimultaneousTCP}. Alternatively, clients can utilize some other common-used schemes (e.g., TURN relaying) for NAT punch-through, which will not be affected~\cite{P2P-NAT}.}
% As most routers are Linux based, and they rely on the Linux iptables module when recording NAT mappings. It's easy for these devices to adopt this strategy by setting the random option of NAT flags, i.e., \textit{nft add rule nat postrouting masquerade random,persistent}\cite{}.
%https://wiki.nftables.org/wiki-nftables/index.php/Performing_Network_Address_Translation_(NAT)

\noindent \textbf{Reverse Path Validation}. Another effective measure to prevent the attack is to adopt the RFC 3704 recommendation, which suggests using the strict mode to filter out forged packets. 
% In essence, packets from the LAN to the router cannot be specified with an outside server's source IP address and should be dropped by the router.
In our test, routers from ASUS, Netgear, ZTE, Aruba, Cisco Meraki, and certain models of TP-LINK, Mercury, and Huawei take this recommendation by default, thus defending against our attack. However, this strategy may introduce additional performance overhead and potentially impact the reliability of networking for certain applications (e.g., OpenVPN running on the router may be affected as the reverse path validation may interfere with packet delivery \cite{openvpn}). 
% https://forum.openwrt.org/t/why-is-rp-filter-disabled-by-default/55434

\noindent \textbf{TCP Window Tracking}. 
As a middle device between the internal clients and outside servers, the router has to keep the necessary information about connections. However, most routers have disabled TCP window tracking for performance reasons. Nevertheless, we find that a simple TCP \texttt{RST} packet can be abused to clear the NAT mapping and be leveraged to launch our attack. In this way, we believe it essential to strictly check the sequence and acknowledgment numbers for received packets. The OpenWrt community has implemented this mitigation as they believe the performance impact should not matter anymore on any currently supported hardware.

%-------------------------------------------------------------------------------
\section{Related Work}\label{sec:relatedwork}
%In this section, we survey the previous related works from the following two aspects, i.e., breaking Wi-Fi security and hijacking network traffic.

%\noindent \textbf{Traffic Manipulating.}
%
Traffic hijacking has been widely studied, and lots of attacks have been proposed. Vulnerabilities that lead to traffic hijacking may exist in protocols at all levels of the TCP/IP protocol stack. For instance, in the same LAN, an attacker can exploit the vulnerability of the ARP protocol to hijack network traffic by sending fake ARP packets and compromising the victim device's ARP cache, which allows the attacker to intercept, modify, or even discard the traffic of victims, thus hijacking the victims' traffic completely \cite{arp-attack}.

At the IP layer, attackers may leverage the ICMP redirect mechanism to hijack victims' traffic by placing themselves in the man-in-the-middle position \cite{Sans-ICMP,icmp-wild}. Recently, Feng \textit{et al.} developed a new method to circumvent the ICMP redirect legitimacy checks in Wi-Fi networks and presented an attack to evade the security mechanisms of WPAs~\cite{feng23mitm}. However, the attack targets out-of-date systems (e.g., iOS 1-8, Android before 10.0) except for the latest versions of Linux and FreeBSD. Besides, they also showed that off-path attackers from the Internet could trick public servers into redirecting their traffic to neighboring hosts with forged ICMP redirect messages, thus causing a DoS attack \cite{secFeng}.

DNS cache poisoning attacks can also be abused to hijack traffic. In the same LAN, Herzberg \textit{et al.} proposed three methods to circumvent source port randomization, which leverages the port allocation strategies used by NAT devices \cite{tpds-Herzberg}. Zheng \textit{et al.} developed an attack targeting DNS forwarders (e.g., home routers) by forcing fragmentation using attacker-owned authoritative name servers \cite{DNS-Zheng-2020}. Man \textit{et al.} proposed that a purely off-path attacker from the Internet can exploit the side channel in ICMP rate limit or the limited space for storing the next hop exception cache to infer the source ports of DNS requests and poison DNS caches maliciously \cite{DNS-Man-2020,DNS-Man-2021}.

To hijack TCP connections so as to inject forged TCP segments into the target connection or terminate it, attackers mainly rely on various side-channel vulnerabilities. Cao \textit{et al.} demonstrated that a global shared variable used in the challenge ACK mechanism could be abused for an off-path attacker to manipulate the victim TCP traffic \cite{cao2016off}. Chen \textit{et al.} showed that a timing side channel that exists in half-duplex IEEE 802.11 or Wi-Fi technology \cite{chen2018off} and Feng \textit{et al.} discovered a side channel in the mixed IPID assignment \cite{ccsfeng,feng2022tonoff}, which can also be exploited to manipulate TCP traffic by off-path attackers. Tolley \textit{et al.} demonstrated that blind in/on-path attackers could learn the virtual IP of a host behind a VPN and hijack TCP connections supposedly protected by the tunnel \cite{Tolley21VPN}.
Besides, Schepers \textit{et al.} discovered that modern operating systems fail to manage the security context of their transmit queues securely, thereby allowing a malicious attacker to intercept frames in Wi-Fi networks, thus hijacking TCP connections or intercepting client and web traffic \cite{schepers2023framing}.

% Song \textit{et al.} proposed to poison OSPF routing table \cite{song2017novel} and Cho \textit{et al.} proposed to announce anomalous BGP messages \cite{cho2019bgp} to launch routing hijacking attacks in control planes, which also allow attackers to manipulate network traffic. By utilizing the mechanism of IP fragmentation, attackers may inject fake IP fragments into the victim connection and poison the targeted traffic once the fake fragments bypass the security checks and are reassembled with the benign ones \cite{gilad2011fragmentation,gilad2013fragmentation,herzberg2013fragmentation,brandt2018domain}.

Fortunately, most of the prior vulnerabilities have already been addressed \cite{arp-1,cao2016off,feng2022tonoff}, and the security community has developed corresponding defense measures against these attacks \cite{rfc1981,RFC1191,rfc8205}. 
However, we present a new type of TCP traffic hijacking attack leveraging the vulnerabilities in routers, which can circumvent traditional defenses against TCP traffic hijacking attacks and lead to new challenges for the security communities.

\section{Conclusion}\label{sec:conclusion}

In this paper, we uncover a new off-path TCP hijacking attack in the Wi-Fi networks that leverages vulnerable routers. We find that a malicious insider can abuse the NAT port preservation strategy and insufficient reverse path validation strategy of the router to infer the existence of TCP connections from the LAN to a remote server and then obtain the sequence and acknowledgment numbers by manipulating the state of NAT mappings with forged reset packets due to the vulnerable routers disabling TCP window tracking strategy. We confirm the vulnerability in a wide range of routers from different manufacturers and evaluate the new attack in different scenarios, such as SSH DoS, FTP hijacking, and HTTP injection in various Wi-Fi networks. Finally, we suggest countermeasures, report the vulnerabilities to the affected manufacturers, and have received positive acknowledgments.


\section*{Acknowledgment}

We thank our shepherd and anonymous reviewers for their
thoughtful comments. This work was in part supported by China National Funds for Distinguished Young Scientists under No.61825204, NSFC under No.62132011, and Beijing Outstanding Young Scientist Program under No.BJJWZYJH01201910003011. Ke Xu is the corresponding author.



%\bibliographystyle{IEEEtranS}


% \end{thebibliography}

\bibliographystyle{IEEEtranS}
\bibliography{reference}

\appendix

\subsection{Experimental Results of Probing the Router's External IP Address}\label{appendix:probe-IP}

% The detailed information of the 30 Wi-Fi networks is shown in Table~\ref{measurement-results}.

\begin{figure}[h]
	\begin{center}
            \vspace{7pt}	
		\includegraphics[width=\linewidth]{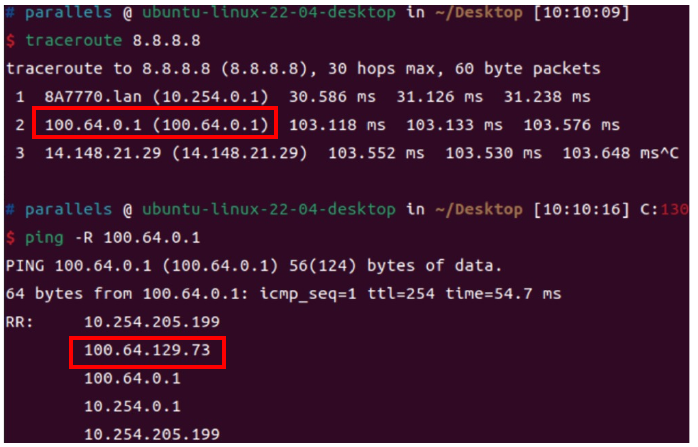}
		% \vspace{2mm}
		\caption{The snapshot of the probing the external IP address of the router.}
		\label{fig:probeExternalIP}
	\end{center}
	% \vspace{-4mm}
\end{figure}

Figure~\ref{fig:probeExternalIP} show the snapshot of the method to probe the external IP address of the router. The attacker, using a laptop with Ubuntu 22.04, has connected to the Wi-Fi network whose router vendor is Wimaster and whose gateway IP is 10.254.0.1. And it has been assigned with the private IP address of 10.254.205.199. Firstly, it gets the gateways along the way to 8.8.8.8 through Traceroute and finds that the second gateway's IP is 100.64.0.1, which is a carried-grade IP address~\cite{imc16_cgn}. Secondly, it pings the second gateway (100.64.0.1) with the \textit{RECORD\_ROUTE} option. The result shows the routes passed, and then the attacker can get the external IP address of the router (i.e., 100.64.129.73).

\subsection{Full List of Tested Routers}\label{appendix:all_routers}

The detailed information of the 67 tested routers is shown in Table~\ref{all-routers}.

\begin{table*}[t]
\vspace{-11mm}
\small
\renewcommand\arraystretch{1.1}
\setlength\tabcolsep{1pt}
\centering
\caption{All Tested Routers}
\label{all-routers}
%\resizebox{\linewidth}{!}{%
\begin{threeparttable}

\scalebox{0.75}{
\setlength{\tabcolsep}{2mm}{
\begin{tabular}{l|l|ccc|cccc|c} 
\bottomrule

\textbf{No.} & \textbf{Router Model} & \textbf{Vendor } & \textbf{OS} & \textbf{Generation }   
& \textbf{\tabincell{c}{Port \\Preservation}}
& \textbf{\tabincell{c}{Reverse-path \\Validation \\ Disabled}}
& \textbf{\tabincell{c}{TCP Window \\Tracking \\ Disabled}}
& \textbf{\tabincell{c}{TCP Close \\Timeout \\(second)}}
& \textbf{Vulnerable} \\ 

\hline

1 & TL-XDR6020 & TP-Link & Linux-based & Wi-Fi 6  & \blackcheck & \blackcheck & \blackcheck  &1 & \redcheck \\
2 & TL-R473GP-AC$^*$ & TP-Link & Linux-based & - & \blackcheck & \blackcheck & \blackcheck &10 & \redcheck \\ 
3 & TL-R4239GP$^*$ & TP-Link & Linux-based & - & \blackcheck & \blackcheck & \blackcheck &1 & \redcheck \\ 
4 & TL-WAR1200L & TP-Link & Linux-based & Wi-Fi 5 & \blackcheck & \blackcheck & \blackcheck &1 & \redcheck \\
5 & TL-R476G & TP-Link & Linux-based & Wi-Fi 5 & \blackcheck & \blackcheck & \blackcheck &1 & \redcheck \\
6 & TL-WDR7620 & TP-Link & Vxworks-based & Wi-Fi 5  & \blackcheck & \blackcross & \blackcheck &1  & \greencross \\
7 & TL-WR886N & TP-Link & Vxworks-based & Wi-Fi 4  & \blackcheck & \blackcross & \blackcheck &1  & \greencross \\ 
\hline

\rowcolor[rgb]{0.871,0.871,0.871} 8 & AX3 Pro & Huawei & EMUI (Linux-based) & Wi-Fi 6 & \blackcheck & \blackcheck & \blackcheck & 10 & \redcheck \\
\rowcolor[rgb]{0.871,0.871,0.871} 9 & AR6140E-9G-2AC$^*$ & Huawei & VRP (Linux-based) & -    & \blackcross & \blackcross & \blackcheck &10 & \greencross \\
\rowcolor[rgb]{0.871,0.871,0.871} 10 & TC7102 & Huawei & EMUI (Linux-based) & Wi-Fi 6 & \blackcheck & \blackcheck & \blackcheck &10 & \redcheck \\
\rowcolor[rgb]{0.871,0.871,0.871} 11 & TC7001 & Huawei & EMUI (Linux-based) & Wi-Fi 6 & \blackcheck & \blackcheck & \blackcheck &10 & \redcheck \\
\rowcolor[rgb]{0.871,0.871,0.871} 12 & Q2S & Huawei & EMUI (Linux-based) & Wi-Fi 5 & \blackcheck & \blackcheck & \blackcheck &10 & \redcheck \\
\rowcolor[rgb]{0.871,0.871,0.871} 13 & WS5200 & Huawei & EMUI (Linux-based) & Wi-Fi 5 & \blackcheck & \blackcheck & \blackcheck &10 & \redcheck \\
\hline

14 & T6M  & 360 & 360OS(Linux-based) & Wi-Fi 6  & \blackcheck & \blackcheck & \blackcheck &10 & \redcheck \\
15 & V6G  & 360 & 360OS(Linux-based) & Wi-Fi 6  & \blackcheck & \blackcheck & \blackcheck &1 & \redcheck \\
16 & T5G  & 360 & 360OS(Linux-based) & Wi-Fi 5  & \blackcheck & \blackcheck & \blackcheck &10 & \redcheck \\
17 & P1  & 360 & 360OS(Linux-based) & Wi-Fi 4  & \blackcheck & \blackcheck & \blackcheck &1 & \redcheck \\
\hline

\rowcolor[rgb]{0.871,0.871,0.871} 18 & Magic R100  & H3C & Comware(Linux-based) & Wi-Fi 5  & \blackcheck & \blackcheck & \blackcheck &10 & \redcheck \\
\rowcolor[rgb]{0.871,0.871,0.871} 19 & Magic R365  & H3C & Comware(Linux-based) & Wi-Fi 5  & \blackcheck & \blackcheck & \blackcheck &10 & \redcheck \\
\rowcolor[rgb]{0.871,0.871,0.871} 20 & Magic R2+  & H3C & Comware(Linux-based) & Wi-Fi 5  & \blackcheck & \blackcheck & \blackcheck &10 & \redcheck \\
\hline

21 & W30E  & Tenda & Linux-based & Wi-Fi 6 & \blackcheck & \blackcheck & \blackcheck & 1 & \redcheck \\
22 & EM12  & Tenda & Linux-based & Wi-Fi 6 & \blackcheck & \blackcheck & \blackcheck & 1 & \redcheck \\
\hline

\rowcolor[rgb]{0.871,0.871,0.871} 23 & RAX1800Z  & China Mobile & AOS(Linux-based) & Wi-Fi 6  & \blackcheck & \blackcheck & \blackcheck & 10 & \redcheck \\
\hline

24 & EG105G$^*$ & Ruijie & RGOS(Linux-based) & - & \blackcheck & \blackcheck & \blackcheck & 1 & \redcheck \\
25 & EG105G-V2$^*$ & Ruijie & RGOS(Linux-based) & - & \blackcheck & \blackcheck & \blackcheck & 1 & \redcheck \\
26 & EG210G-P$^*$ & Ruijie & RGOS(Linux-based) & -   & \blackcheck & \blackcheck & \blackcheck & 1 & \redcheck \\
27 & NBR$^*$ & Ruijie & RGOS(Linux-based) & - & \blackcheck & \blackcheck & \blackcheck & 1 & \redcheck \\
28 & X32 Pro & Ruijie & RGOS(Linux-based) & Wi-Fi 6  & \blackcheck & \blackcheck & \blackcheck  &1 & \redcheck \\
\hline
% openwrt
\rowcolor[rgb]{0.871,0.871,0.871} 29 & Redmi RA81  & Xiaomi & MiWiFi(Linux-based) & Wi-Fi 6  & \blackcheck & \blackcheck & \blackcheck &1 & \redcheck \\
\rowcolor[rgb]{0.871,0.871,0.871} 30 & Redmi RA67  & Xiaomi & MiWiFi(Linux-based) & Wi-Fi 6  & \blackcheck & \blackcheck & \blackcheck &1 & \redcheck \\
\rowcolor[rgb]{0.871,0.871,0.871} 31 & R3L  & Xiaomi & MiWiFi(Linux-based) & Wi-Fi 6 & \blackcheck & \blackcheck & \blackcheck &10 & \redcheck \\
\rowcolor[rgb]{0.871,0.871,0.871} 32 & R3G  & Xiaomi & MiWiFi(Linux-based) & Wi-Fi 5 & \blackcheck & \blackcheck & \blackcheck &10 & \redcheck \\
\rowcolor[rgb]{0.871,0.871,0.871} 33 & CR6609  & Xiaomi & MiWiFi(Linux-based) & Wi-Fi 6 & \blackcheck & \blackcheck & \blackcheck &10 & \redcheck \\
\hline

34 & MW300R & Mercury & Vxworks-based & Wi-Fi 4   & \blackcheck & \blackcross & \blackcheck &1 & \greencross \\
35 & X30G & Mercury & Linux-based & Wi-Fi 6  & \blackcheck & \blackcheck & \blackcheck &1 & \redcheck \\
36 & D121G & Mercury & Vxworks-based & Wi-Fi 5  & \blackcheck & \blackcross & \blackcheck &1 & \greencross \\
37 & YR1900MG & Mercury & Vxworks-based & Wi-Fi 5  & \blackcheck & \blackcross & \blackcheck &1 & \greencross \\
38 & YR1800XG & Mercury & Linux-based & Wi-Fi 6  & \blackcheck & \blackcheck & \blackcheck &1 & \redcheck \\
\hline

\rowcolor[rgb]{0.871,0.871,0.871} 39 & RAX20  & Netgear & DumaOS(Linux-based) & Wi-Fi 6  & \blackcheck & \blackcross & \blackcheck &10 & \greencross \\
\rowcolor[rgb]{0.871,0.871,0.871} 40 & RAX50  & Netgear & DumaOS(Linux-based) & Wi-Fi 6  & \blackcheck & \blackcross & \blackcheck &10 & \greencross \\
\hline

41 & RT-AX57 & ASUS & AsusWrt(Linux-based) & Wi-Fi 6   & \blackcheck & \blackcross & \blackcheck &10 & \greencross \\
42 & RT-AX89X & ASUS & AsusWrt(Linux-based) & Wi-Fi 6   & \blackcheck & \blackcross & \blackcheck &10 & \greencross \\
\hline

\rowcolor[rgb]{0.871,0.871,0.871} 43 & E5600 & Linksys & Linux-based & Wi-Fi 6   & \blackcheck & \blackcheck & \blackcheck &10 & \redcheck \\
\rowcolor[rgb]{0.871,0.871,0.871} 44 & E9450 & Linksys & Linux-based & Wi-Fi 6  & \blackcheck & \blackcheck & \blackcheck &10 & \redcheck \\
\hline

45 & QUANTUM D2G & Wavlink & Linux-based & Wi-Fi 5 & \blackcheck & \blackcheck & \blackcheck &10 & \redcheck \\
\hline

% \rowcolor[rgb]{0.871,0.871,0.871} 46 & CF-616AC & Comfast & OrangeOS(OpenWrt-based) & Wi-Fi 5  & \blackcheck & \blackcheck & \blackcheck &10 & \redcheck \\
% \hline
\rowcolor[rgb]{0.871,0.871,0.871} 46 & CF-616AC & Comfast & OrangeOS(Linux-based) & Wi-Fi 5  & \blackcheck & \blackcheck & \blackcheck &10 & \redcheck \\
\hline

47 & DI-7003GV2$^*$ & D-Link & Linux-based & -  & \blackcheck & \blackcheck & \blackcheck &1 & \redcheck \\
\hline

\rowcolor[rgb]{0.871,0.871,0.871} 48 & E3630  & ZTE & ZXR10ROS(Linux-based) & Wi-Fi 6 & \blackcheck & \blackcross & \blackcheck &10 & \greencross \\
\rowcolor[rgb]{0.871,0.871,0.871} 49 & AX3000  & ZTE & ZXR10ROS(Linux-based) & Wi-Fi 6 & \blackcheck & \blackcross & \blackcheck &10 & \greencross \\
\hline

50 & M80$^*$ & IP-COM  & Linux-based & -  & \blackcheck & \blackcheck & \blackcheck &1 & \redcheck \\
\hline

\rowcolor[rgb]{0.871,0.871,0.871} 51 & SK-WR6640X  & Skyworth  & Linux-based & Wi-Fi 6   & \blackcheck & \blackcheck & \blackcheck &10 & \redcheck \\
\hline

52 & VX3000 & Volans   & Linux-based & Wi-Fi 6   & \blackcheck & \blackcheck & \blackcheck &1 & \redcheck \\
53 & VE5200G$^*$ & Volans   & Linux-based & - & \blackcheck & \blackcheck & \blackcheck &1 & \redcheck \\
\hline

\rowcolor[rgb]{0.871,0.871,0.871} 54 & MG1200AC & Netcore  & NOS(Linux-based) & Wi-Fi 5  & \blackcheck & \blackcheck & \blackcheck &1 & \redcheck \\
\rowcolor[rgb]{0.871,0.871,0.871} 55 & NBR1009GPE & Netcore  & NOS(Linux-based) & -  & \blackcheck & \blackcheck & \blackcheck &1 & \redcheck \\
\hline

% 56 & Wimaster-mini$^*$ & Wimaster  & OpenWrt-based & - & \blackcheck & \blackcheck & \blackcheck & 10 & \redcheck \\
% 57 & Wimaster$^*$ & Wimaster  & OpenWrt-based & - & \blackcheck & \blackcheck & \blackcheck & 10 & \redcheck \\
% \hline
56 & Wimaster-mini$^*$ & Wimaster  & Linux-based & - & \blackcheck & \blackcheck & \blackcheck & 10 & \redcheck \\
57 & Wimaster$^*$ & Wimaster  & Linux-based & - & \blackcheck & \blackcheck & \blackcheck & 10 & \redcheck \\
\hline

% \rowcolor[rgb]{0.871,0.871,0.871} 58 & IK-Q90 & iKuai  & iKuaiOS(OpenWrt-based) & Wi-Fi 6  & \blackcheck & \blackcheck & \blackcheck &10 & \redcheck \\
% \rowcolor[rgb]{0.871,0.871,0.871} 59 & IK-Enterprise$^*$ & iKuai  & iKuaiOS(OpenWrt-based) & -  & \blackcheck & \blackcheck & \blackcheck &10 & \redcheck \\
% \hline
\rowcolor[rgb]{0.871,0.871,0.871} 58 & IK-Q90 & iKuai  & iKuaiOS(Linux-based) & Wi-Fi 6  & \blackcheck & \blackcheck & \blackcheck &10 & \redcheck \\
\rowcolor[rgb]{0.871,0.871,0.871} 59 & IK-Enterprise$^*$ & iKuai  & iKuaiOS(Linux-based) & -  & \blackcheck & \blackcheck & \blackcheck &10 & \redcheck \\
\hline

60 & Instant On AP22 & Aruba & ArubaOS(Linux-based) & Wi-Fi 6 & \blackcheck & \blackcross & \blackcheck & 10 & \greencross \\
\hline

\rowcolor[rgb]{0.871,0.871,0.871} 61 &  EdgeRouter X$^*$ & Ubiquiti  & Linux-based & - & \blackcheck & \blackcheck & \blackcheck &10 & \redcheck \\
\hline

62 & AX1800 & JdCloud  & Linux-based & Wi-Fi 6 & \blackcheck & \blackcheck & \blackcheck & 10 & \redcheck \\ 
\hline

\rowcolor[rgb]{0.871,0.871,0.871} 63 & Cisco Meraki 64$^*$ & Cisco Meraki & Linux-based & - & \blackcheck & \blackcross & \blackcross & - & \greencross \\
\hline

64 & eero pro & Amazon & Linux-based & Wi-Fi 5 & \blackcheck & \blackcheck & \blackcheck & 10 & \redcheck \\
\hline

\rowcolor[rgb]{0.871,0.871,0.871} 65 & Google Wi-Fi & Google  & ChromeOS(Linux-based) & Wi-Fi 5 & \blackcheck & \blackcheck & \blackcheck & 10 & \redcheck \\ 
\hline

66 & GL-MT3000 & GL.iNet  & Linux-based & Wi-Fi 6 & \blackcheck & \blackcheck & \blackcheck & 10 & \redcheck \\
\hline

\rowcolor[rgb]{0.871,0.871,0.871} 67 & pfSense 2.7.0$^*$ & pfSense & FreeBSD-based & - & \blackcross & \blackcross & \blackcheck & 90 & \greencross \\

\toprule
\end{tabular}}}
   \begin{tablenotes}
       \footnotesize
       \item[] \blackcheck means that the router is satisfied with the condition, and \blackcross means that the router is dissatisfied with the condition.
       \item[] \redcheck means that the router is vulnerable to our attack, and \greencross means that the router is immune to our attack.
       \item[] * means that the model is an enterprise router which does not support Wi-Fi by itself and needs to work together with wireless access points.
    \end{tablenotes}
\end{threeparttable}
%}
% \vspace{-5mm}
\end{table*}

\end{document}